\documentclass{aa}
\usepackage[T1]{fontenc}
\usepackage[utf8]{inputenc}

\usepackage{newtxtext}  
\usepackage[varg]{newtxmath}  
\usepackage{amsmath}    
\usepackage{amssymb}    
\usepackage{nicefrac}   
\usepackage{textcomp}
\usepackage{url}
\usepackage{bm}         
\usepackage{microtype}
\usepackage{multirow}

\usepackage[exponent-product=\cdot
            ,tight-spacing=true
            ,range-phrase={\text{--}}
            ,range-units=single
            ,list-units=single
            ,angle-symbol-over-decimal=true
            ,detect-weight=true
            ,list-pair-separator= {\text{, and }}
            ,list-final-separator= {\text{, and }}
            ]{siunitx}
\DeclareSIUnit\au{au}
\DeclareSIUnit\pc{pc}
\DeclareSIUnit\jy{Jy}
\DeclareSIUnit\msun{M\ensuremath{_{\sun}}}
\DeclareSIUnit\lsun{L\ensuremath{_{\sun}}}

\usepackage{graphicx}
    \DeclareGraphicsExtensions{.pdf}
    \setkeys{Gin}{width=\linewidth}
    \setkeys{Gin}{height=\textheight}
    \setkeys{Gin}{keepaspectratio=true}
    \graphicspath{{./plots/}}
\usepackage{float}
\usepackage{tikz}
\usepackage{tikzscale}

\usepackage{hyperref}
\makeatletter
\renewcommand*\aa@pageof{, page \thepage{} of \pageref*{LastPage}}
\makeatother

\usepackage{prettyref}  
\newrefformat{fig}{\text{Fig.~\ref{#1}}}
\newrefformat{eq}{\text{Eq.~\ref{#1}}}
\newrefformat{tab}{\text{Table~\ref{#1}}}
\newrefformat{sec}{\text{Sect.~\ref{#1}}}
\newrefformat{enu}{\text{~\ref{#1}}}

\newcommand{\pref}[1]{\prettyref{#1}}

\renewcommand{\href}{\ensuremath{h_{\text{ref}}}}

\title{Self-scattering in protoplanetary disks with dust settling}
\titlerunning{Self-scattering of settled dust grains}

\author{R. Brunngräber \and S. Wolf}
\institute{Institut für Theoretische Physik und Astrophysik, Christian-Albrechts-Universität zu Kiel, Leibnizstr. 15, 24118 Kiel, Germany\\\email{rbrunngraeber@astrophysik.uni-kiel.de}}

\date{Received / Accepted}

\abstract{Scattering of re-emitted flux is considered to be at least partially responsible for the observed polarisation in the (sub-)millimetre wavelength range of several protoplanetary disks. Although the degree of polarisation produced by scattering is highly dependent on the dust model, early studies investigating this mechanism relied on the assumption of single grain sizes and simple density distribution of the dust. However, in the dense inner regions where this mechanism is usually most efficient, the existence of dust grains with sizes ranging from nanometres to millimetres has been confirmed. Additionally, the presence of gas forces larger grains to migrate vertically towards the disk midplane, introducing a dust segregation in the vertical direction. Using polarisation radiative transfer simulations, we analyse the dependence of the resulting scattered light polarisation at 350\,\textmu m, 850\,\textmu m, 1.3\,mm, and 2\,mm on various parameters describing protoplanetary disks, including the effect of dust grain settling. We find that the different disk parameters change the degree of polarisation mostly by affecting the anisotropy of the radiation field, the optical depth, or both. It is therefore very challenging to deduce certain disk parameter values directly from polarisation measurements alone. However, assuming a high dust albedo, it is possible to trace the transition from optically thick to optically thin disk regions. The degree of polarisation in most of the considered disk configurations is lower than what is found observationally, implying the necessity to revisit models that describe the dust properties and disk structure.}

\keywords{Radiative transfer -- Protoplanetary disks -- Polarization -- Radiation mechanisms: thermal -- Scattering}
\bibpunct{(}{)}{;}{a}{}{,}

\begin{document}
\maketitle



\section{Introduction}
    Observations of polarised light are key to understanding the dust properties in various objects related to star and planet formation, such as Bok globules, filaments, and protoplanetary disks. The light coming from these objects is mostly polarised due to dichroic emission and absorption \citep{andersson-lazarian-vaillancourt-2015} and scattering \citep{kataoka-et-al-2015,weintraub-goodman-akeson-2000}. Dichroic emission and absorption plays an important role in determining the structure and strength of magnetic fields if it is assumed that non-spherical dust grains align with their major axis perpendicular to the magnetic field lines. The second, complementary process leading to polarisation is scattering. At optical and near-infrared wavelengths, the scattered flux comes originally from the stellar source. In the mid-infrared, and especially at submillimetre (sub-mm) and millimetre (mm) wavelengths, the scattered radiation arises originally from the thermal, re-emitted light of the dust grains themselves, hence the expression 'self-scattering'. Various observations of protoplanetary disks obtained in the (sub-)mm wavelength range performed with the Atacama Large Millimeter/submillimeter Array (ALMA), as well as the \textit{N} band observation of AB\,Aurigae, show polarisation patterns that may be explained by scattering \citep{ohashi-kataoka-2019,dent-et-al-2019,lee-et-al-2018,bacciotti-et-al-2018,ohashi-et-al-2018,hull-et-al-2018,stephens-et-al-2017,li-et-al-2016,stephens-et-al-2014}.

    The scattering properties of grains, and thus the observed polarised fraction and orientation, depend on many parameters, such as grain size, shape, internal structure (compact, porous, layered), chemical composition, observing wavelength, and scattering angle \citep[e.g.][]{kirchschlager-bertrang-2020,brunngraeber-wolf-2019,min-hovenier-dekoter-2005,wolf-voshchinnikov-henning-2002,pendleton-tielens-werner-1990,dolginov-silantev-1976}. Based on these microphysical properties, the net scattered light polarisation of an astrophysical object is determined by the spatial distribution of the number density and the aforementioned properties of the scatterers (i.e. the dust phase). In turn, the observed polarisation, especially at multiple wavelengths, can in principle be used to constrain these parameters. To date, most studies covering self-scattering make use of one or multiple crucial simplifications, for example single grain size instead of a size distribution, a homogeneous mixture of different grain sizes instead of size-dependent distribution, semi-analytic radiative transfer instead of full Monte Carlo simulations, single instead of multiple scattering, or simple density distributions. To fully evaluate the impact of polarisation from scattered re-emission radiation from protoplanetary disks, some of these simplifications may lead to false conclusions if applied to real protoplanetary disk observations. In particular, the number and spatial distribution of grain sizes are expected to have a major impact on the resulting polarisation degree.

    In this paper, we investigate the polarised light due to scattering at far-infrared to mm wavelengths considering a protoplanetary disk with the largest dust grains already settled towards the disk midplane. Different scale heights for different grain sizes are expected due to the size-dependent friction between dust and gas in such a disk. The disk morphology and the grain properties used in this study are introduced in \pref{sec:set-up}. The results of the 3D Monte Carlo radiative-transfer simulations performed with the publicly available code \texttt{POLARIS}\footnote{\url{http://www1.astrophysik.uni-kiel.de/~polaris/index.html}} \citep{reissl-et-al-2016} are shown in \pref{sec:results}. In \pref{sec:discussion}, we discuss the impact of our results on the analysis of polarisation observations and summarise our findings in \pref{sec:summary}.
\section{Model set-up}
\label{sec:set-up}
    The underlying model for our simulations is that of a protoplanetary disk. The volume density distribution of the gas disk is the most widely used model in protoplanetary disk studies \citep[e.g.][]{kama-et-al-2020,brunngraeber-wolf-2019,trapman-et-al-2019,cox-et-al-2017,andrews-et-al-2011}. It is based on the studies by \citet{lynden-bell-pringle-1974}, \citet{kenyon-hartmann-1995}, and \citet{hartmann-et-al-1998} and follows a power law with exponential decay in radial direction and displays a Gaussian distribution in the vertical direction. It is given by
    \begin{equation}
    \label{eq:dens_distro}
        \rho_{\text{gas}}(r,z) = \frac{\Sigma_{\text{gas}}(r)}{\sqrt{2\pi}\,h_{\text{gas}}(r)}\!\cdot\exp{\left[-\frac{1}{2}\left(\frac{z}{h_{\text{gas}}(r)}\right)^2\right]}\ ,
    \end{equation}
    with the vertically integrated surface density
    \begin{equation}
    \label{eq:sigma}
        \Sigma_{\text{gas}}(r) = \sqrt{2\pi}\,\rho_0\,\href{}\cdot\left(\frac{r}{R_0}\right)^{-\gamma}\cdot\exp{\left[-\left(\frac{r}{R_{\text{trunc}}}\right)^{2-\gamma}\right]}
    \end{equation}
    and the scale height
    \begin{equation}
    \label{eq:scale_height}
        h_{\text{gas}}(r) = \href{}\,\left(\frac{r}{R_0}\right)^{\beta}\ ,
    \end{equation}
    where $(r,z)$ are the usual cylindrical coordiantes, $R_0$ is the radius where the gas scale height $h_{\text{gas}}$ reaches the reference scale height \href{}, $R_{\text{trunc}}$ is the radius where the exponential decay becomes more and more important, and $\rho_0$ is the density scaling parameter which is given by the total disk mass.

    In a protoplanetary disk, the dust grains are bound to the movement of the gas, which make up about $\SI{99}{\percent}$ of the total disk mass. The strength of this coupling depends on the Stokes number and thus mostly on the dust grain size \citep{safronov-1969,cuzzi-et-al-1993,testi-et-al-2014}. As a result, the largest grains that exist in the disk are more concentrated in the midplane, whereas smaller grains may extend to the upmost layers of the gas disk. Therefore, the dust density distribution is expected to differ from the gas distribution depending on the grain size. In this study, we make use of the dust settling model presented by \citet{dubrulle-et-al-1995} and \citet{woitke-et-al-2016}:
    \begin{equation*}
        h_{\text{dust}}(r,s) = h_{\text{gas}}(r)\cdot \sqrt{\frac{f(r,s)}{1+f(r,s)}}\ ,
    \end{equation*}
    where the settling function $f(r,s)$ is given by
    \begin{equation}
    \label{eq:settling}
        f(r,s) = \frac{\alpha}{\sqrt{6\pi}}\ \frac{\Sigma_{\text{gas}}(r)}{s\,\varrho_{\text{bulk}}}\ ,
    \end{equation}
    with $s$ being the radius and $\varrho_{\text{bulk}}$ being the bulk density of the homogeneous, compact, spherical dust grains. The strength of the dust settling is parameterised only by $\alpha$, which is equal to the widely used viscosity parameter for so-called $\alpha$-disks \citep{shakura-sunyaev-1973}. The volume and surface density of the dust, $\rho_{\text{dust}}(r,z)$ and $\Sigma_{\text{dust}}(r)$, are calculated via \pref{eq:dens_distro} and \pref{eq:sigma}, respectively, with the according dust scale height $h_{\text{dust}}(r,s)$. The surface density is therefore not affected by the process of dust settling.

    The dust grains are composed of a mixture of \SI{62.5}{\percent} silicate and \SI{37.5}{\percent} graphite (in mass), using the $\frac{1}{3}$--$\frac{2}{3}$ approximation for parallel and perpendicular orientations of the graphite grains \citep{draine-malhotra-1993}. The complex refractive indices were taken from \citet{draine-lee-1984}, \citet{laor-draine-1993}, and \citet{weingartner-draine-2001}.
    The size distribution follows the power law $n(s)\propto s^{q}$ \citep{mathis-et-al-1977} and is commonly used for protoplanetary disk simulations. The number of dust grains in the size interval $[s,s+\text{d}s]$ is given by $n\text{d}s$. The minimum grain size is $s_{\text{min}}=\SI{5}{\nm}$ and the maximum for most disk models is $s_{\text{max}}=\SI{1}{\mm}$; however, we also simulate one disk with a lower maximum grain size of $s_{\text{max}}=\SI{100}{\um}$. In order to account for dust growth and settling, we divide the global size range into ten logarithmically spaced size bins. The relative dust mass of the $i^{\text{th}}$ size bin is determined by integrating the grain size distribution over the entire volume:    %
    \begin{equation*}
        \frac{M_{\text{dust},i}}{M_{\text{dust}}} = \frac{s^{q+4}_{\text{max},i} - s^{q+4}_{\text{min},i}}{s^{q+4}_{\text{max}} - s^{q+4}_{\text{min}}}\ .
    \end{equation*}
    These size bins are used for the calculation of the density distribution. In addition, each bin is sub-divided into 100 bins, also spaced logarithmically, to better account for size-dependent quantities during the simulation; these include absorption and scattering cross-sections and the Müller matrix, which may vary strongly over a very small range in grain size.

    The orientation of the disk is set to face-on to reduce further impacts on the polarisation resulting from geometrical effects, for instance different scattering angles on the far- and near-side of the disk \citep{yang-et-al-2016a,yang-et-al-2017,brunngraeber-wolf-2019}.

\begin{figure}
        \includegraphics[width=0.495\linewidth]{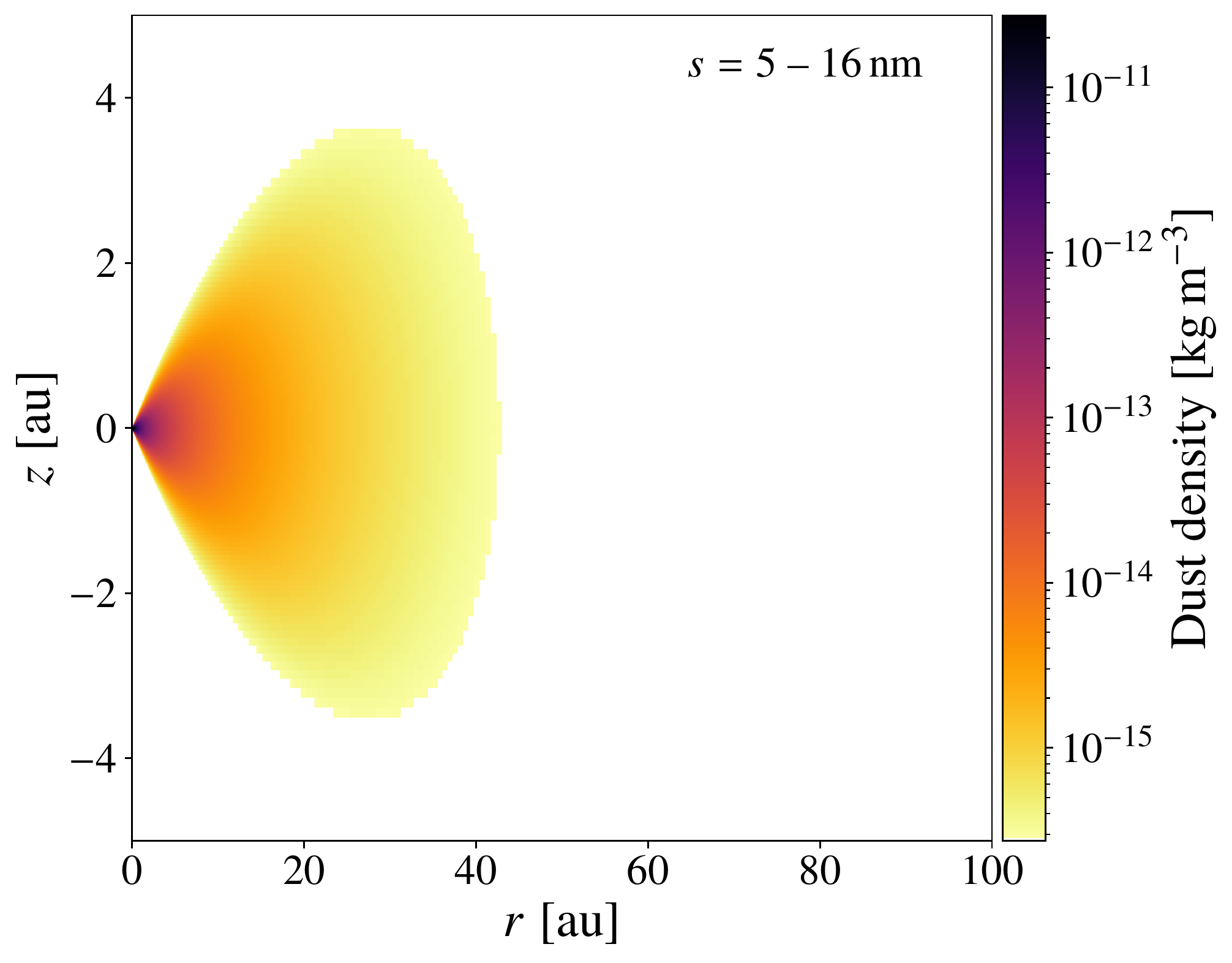}
        \includegraphics[width=0.495\linewidth]{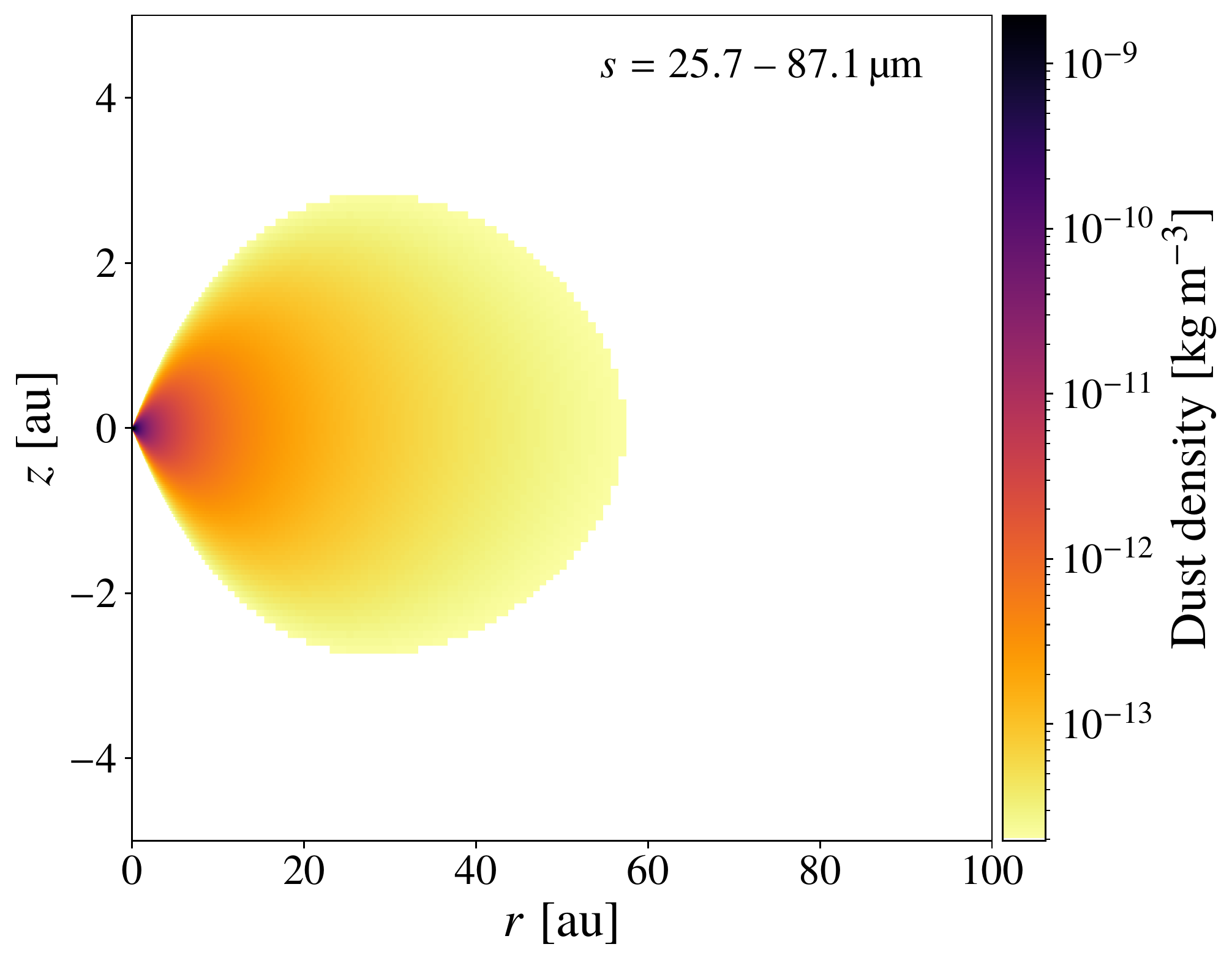}\\
        \includegraphics[width=0.495\linewidth]{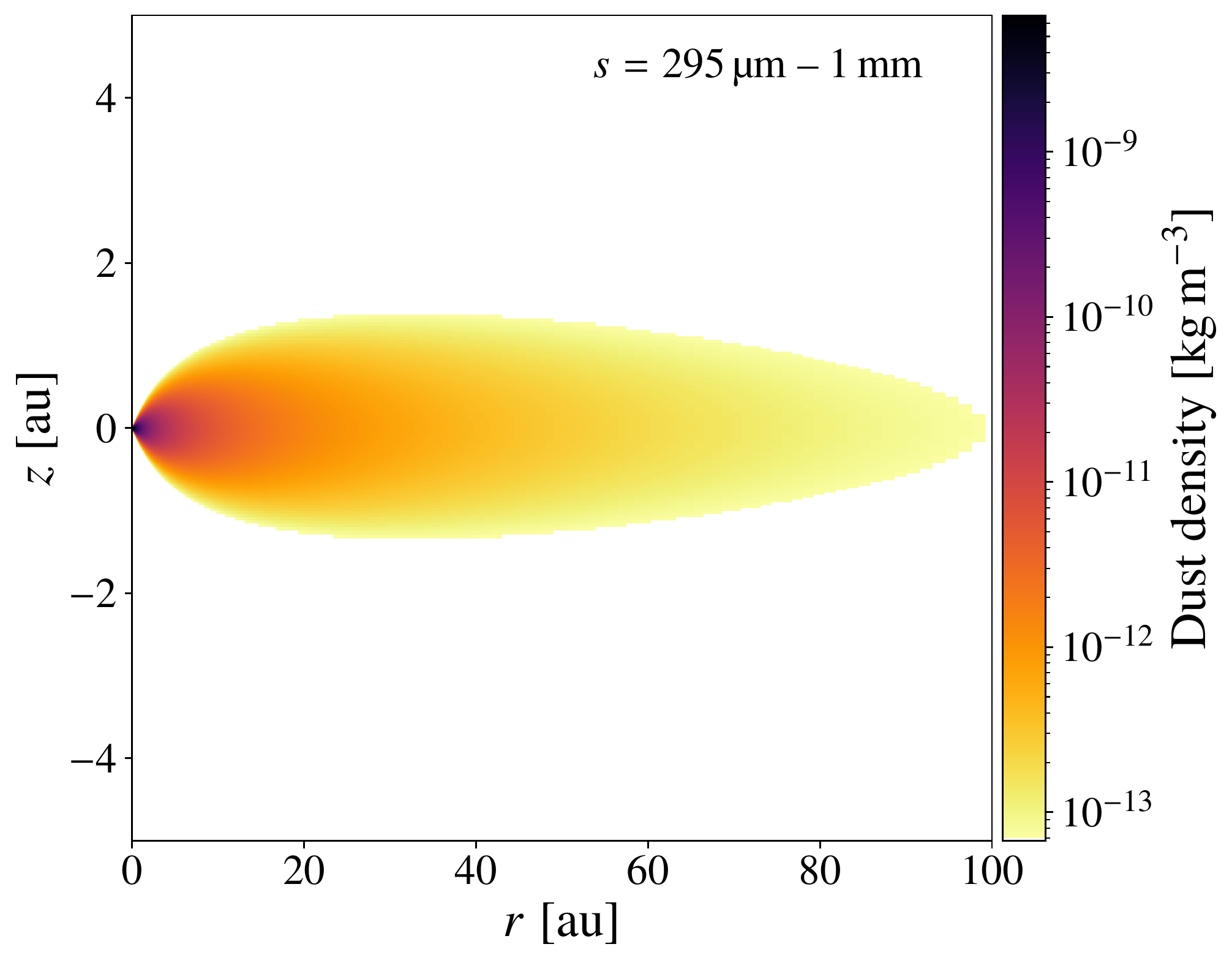}
        \includegraphics[width=0.495\linewidth]{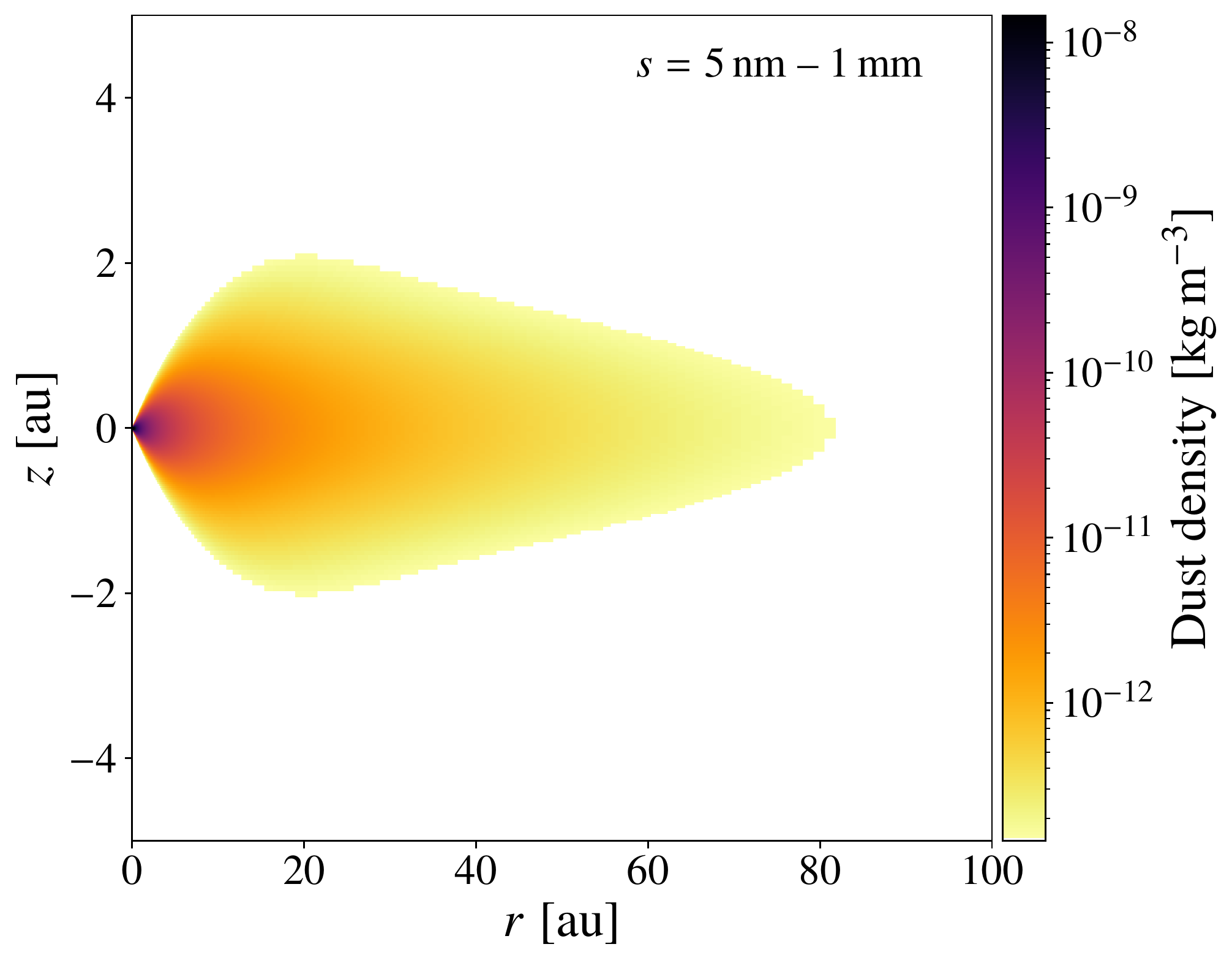}
        \caption{Dust density distribution of the reference disk for three different size bins and for the entire disk, that is, the sum of all grain size bins.}
        \label{fig:dust_distro}
    \end{figure}

\begin{figure}
        \includegraphics[width=0.99\linewidth,height=0.3\textheight]{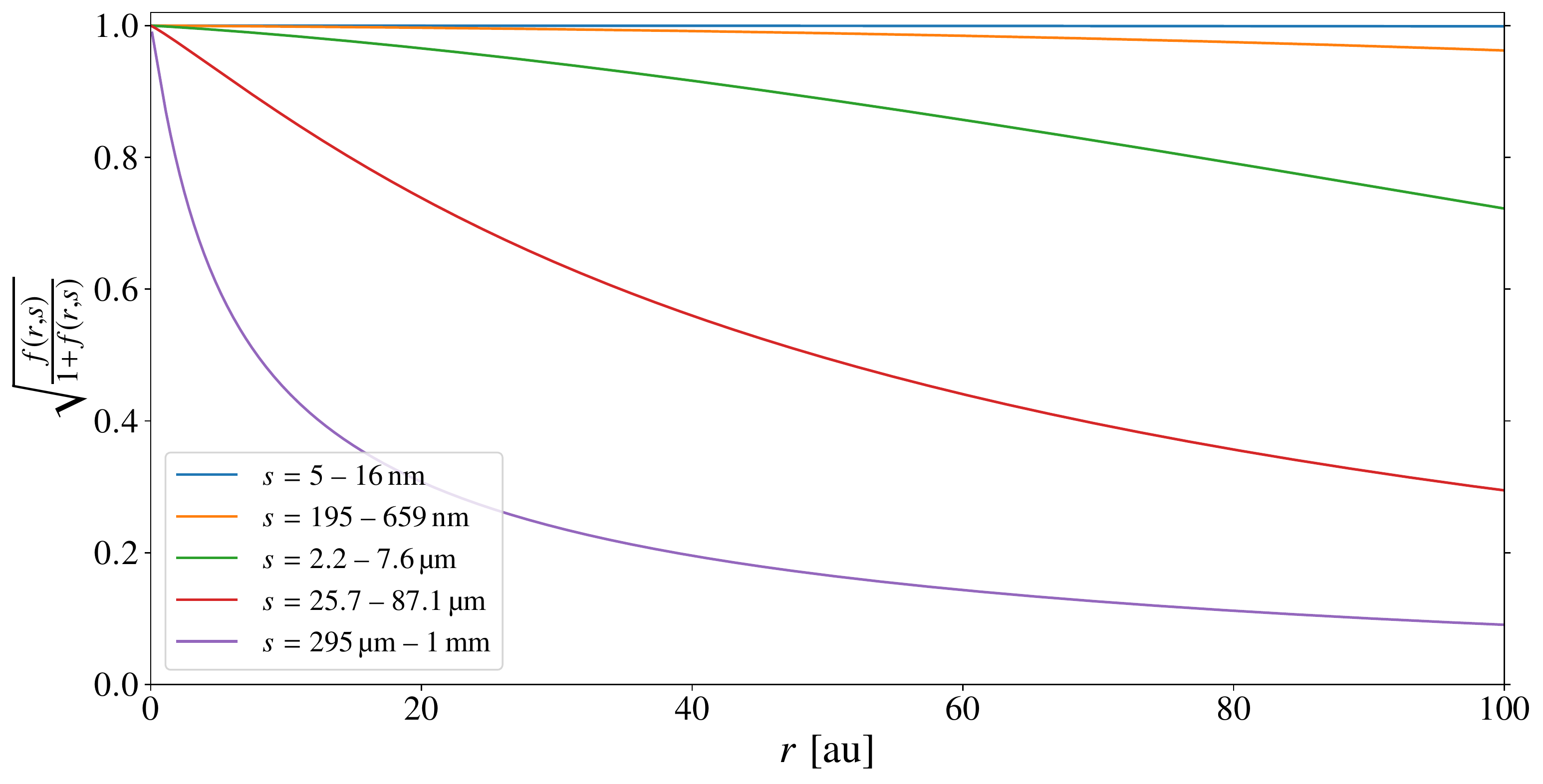}
        \caption{The scaling factor $\sqrt{\frac{f}{1+f}}$ of the dust scale height $h_{\text{dust}}$ for five out of ten grain size bins for the reference disk; see also \pref{eq:settling}.}
        \label{fig:scaling_parameter}
    \end{figure}

    For a compilation of the model parameters, see \pref{tab:disk_param_space}. The dust mass density distribution in the ($r$,$z$)-plane of our reference disk model is shown in \pref{fig:dust_distro}. The upper row and the left-hand panel of the lower row show the density distribution for different grain size bins to illustrate the dust settling. The combination of all grain size bins is shown in the lower-right panel. In addition, \pref{fig:scaling_parameter} shows the dust settling function over the radial distance from the star for some size bins.

    \begin{table}
        \caption{disk and stellar parameter values. If multiple values are given for one parameter, the value in bold face is used for the reference disk model.}
        \label{tab:disk_param_space}
        \centering
        \begin{tabular}{l l l}
            \hline\hline
            \rule{0pt}{2.5ex}Parameter          & Variable                                  & Values    \\[1mm]
            \hline
            \rule{0pt}{2.5ex}Inner radius       & $R_{\text{in}}$ [\si{\au}]                & \num{0.1} \\
            Outer radius                        & $R_{\text{out}}$ [\si{\au}]               & \num{300} \\
            Reference radius                    & $R_{0}$ [\si{\au}]                        & \num{100} \\
            Truncation radius                   & $R_{\text{trunc}}$ [\si{\au}]             & \num{100} \\
            Reference scale height              & \href{} [\si{\au}]                        & \textbf{\num{10}}, \numlist{15; 20; 8} \\
            Density profile                     & $\gamma$                                  & \textbf{\num{1.1}}, \numlist{1.4; 0.8} \\
            Flaring parameter                   & $\beta$                                   & \textbf{\num{1.1}}, \numlist{1.3; 1.5} \\
            Dust mass                           & $M_{\text{dust}}$ [\si{\msun}]            & \textbf{\num{e-4}}, \numlist{e-3; e-5} \\
            Viscosity parameter                 & $\alpha$                                  & \textbf{\num{e-2}}, \numlist{e-3; e-1} \\
            Inclination                         & $i$ [\si{\degree}]                        & \numlist{0} \\
            Stellar luminosity                  & $L_{\star}$ [\si{\lsun}]                  & \num{0.9} \\
            Stellar temperature                 & $T_{\star}$ [\si{\K}]                     & \num{4050} \\[.8ex]
            \multirow{3}{30mm}{Dust composition (mass fraction)} &                          & Silicate: \SI{62.5}{\percent} \\
                                                &                                           & Graphite$_{\perp}$: \SI{25}{\percent} \\
                                                &                                           & Graphite$_{\parallel}$: \SI{12.5}{\percent} \\[.8ex]
            \multirow{3}{*}{Grain bulk density} & \multirow{3}{15mm}{$\varrho_{\text{bulk}}$ [\si{\kg\per\m\cubed}]}    & Silicate: \num{3500} \\
                                                &                                           & Graphite: \num{2250} \\
                                                &                                           & Mixture: $\sim \num{2896.6}$ \\[.8ex]
            Minimum grain size                  & $s_{\text{min}}$ [\si{\nm}]               & \num{5} \\
            Maximum grain size                  & $s_{\text{max}}$ [\si{\mm}]               & \textbf{\num{1}} and \num{0.1} \\
            Size distribution                   & $q$                                       & \textbf{\num{-3.5}}, \numlist{-3.2; -3.8} \\
            Number of size bins                 &                                           & \num{10} \\
            \multirow{2}{*}{Wavelength}         & \multirow{2}{*}{$\lambda$ [\si{\um}]}     & \multirow{2}{30mm}{\numlist{350; 850; 1300; 2000}} \\
                                                &                                           &  \\
            \hline
        \end{tabular}
    \end{table}
\section{Results}
\label{sec:results}
    In this section, we present the influence of the aforementioned disk and dust parameters on the polarisation degree of protoplanetary disks due to the scattered thermal re-emission. Each parameter is investigated separately and the remaining parameters are kept constant. Different parameters affect the degree of polarisation of the disk similarly and can be regrouped by more fundamental effects. First, we discuss the impact of the shape of the radiation field in \pref{sec:res_radfield}. The optical depth, as the most fundamental quantity for the understanding of the general appearance of protoplanetary disks, and the observing wavelength are addressed in \pref{sec:res_tau} and \pref{sec:res_wave}, respectively.

    The results of the radiative transfer simulations are shown as radially averaged profiles of the polarisation degree. These profiles are extracted from spatially resolved synthetic images produced by \texttt{POLARIS}, which in turn are based on full 3D Monte Carlo temperature simulations. The synthetic images consist of 255$\times$255 pixels with a resolution of $\sim\SI{2.35}{\au}\times\SI{2.35}{\au}$. To extract radial profiles, the images are radially binned with respect to the central pixel. Each data point shown in the figures is the weighted arithmetic mean of five such radial bins to increase the signal-to-noise ratio. The resolution is thus decreased to about $\SI{12}{\au}$. Each profile is presented with error bars, which represent the statistical uncertainties that naturally arise from the Monte Carlo method.

    \subsection{Anisotropy of the radiation field}
    \label{sec:res_radfield}
\begin{figure}
            \includegraphics[width=0.495\linewidth]{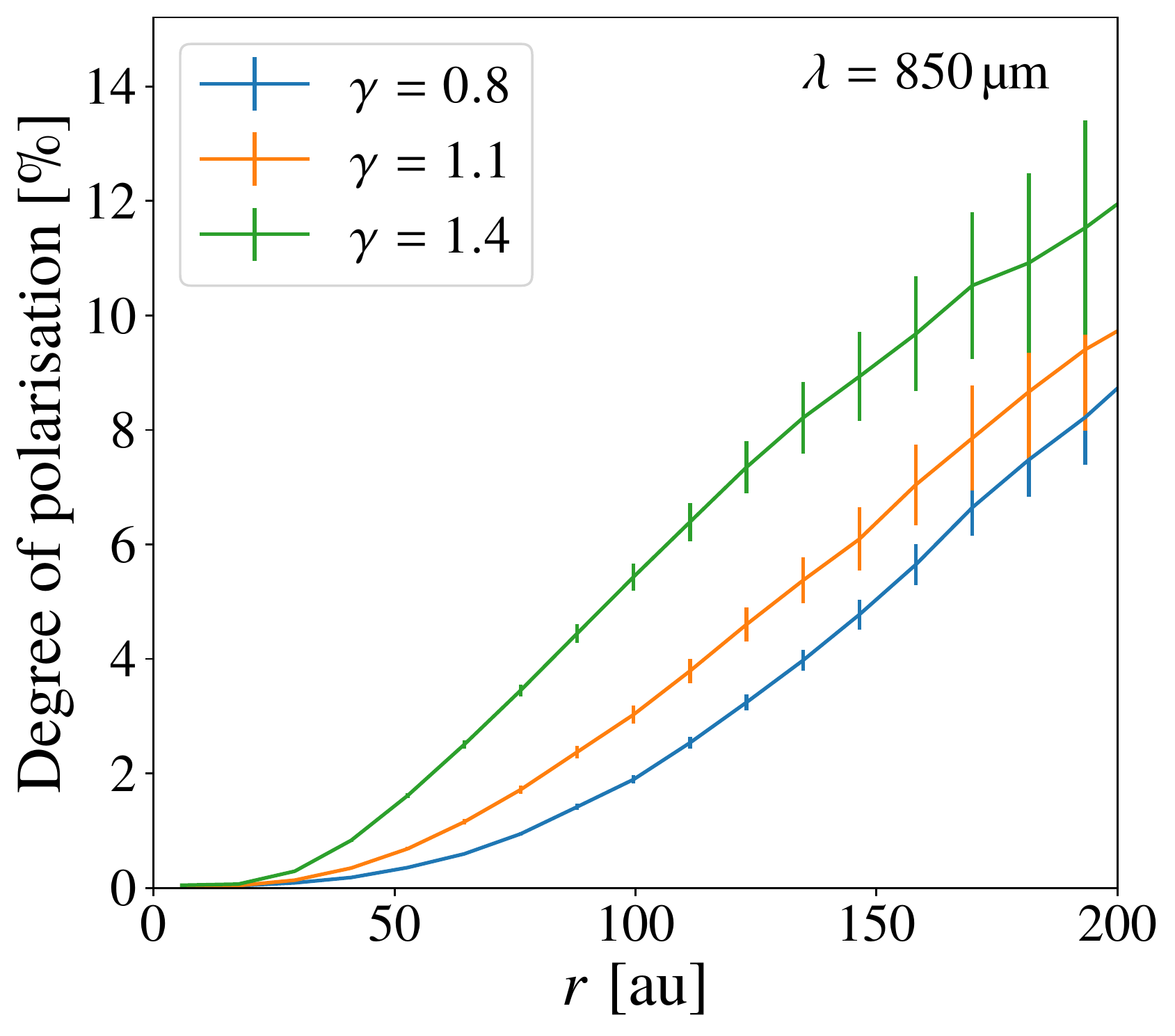}
            \includegraphics[width=0.495\linewidth]{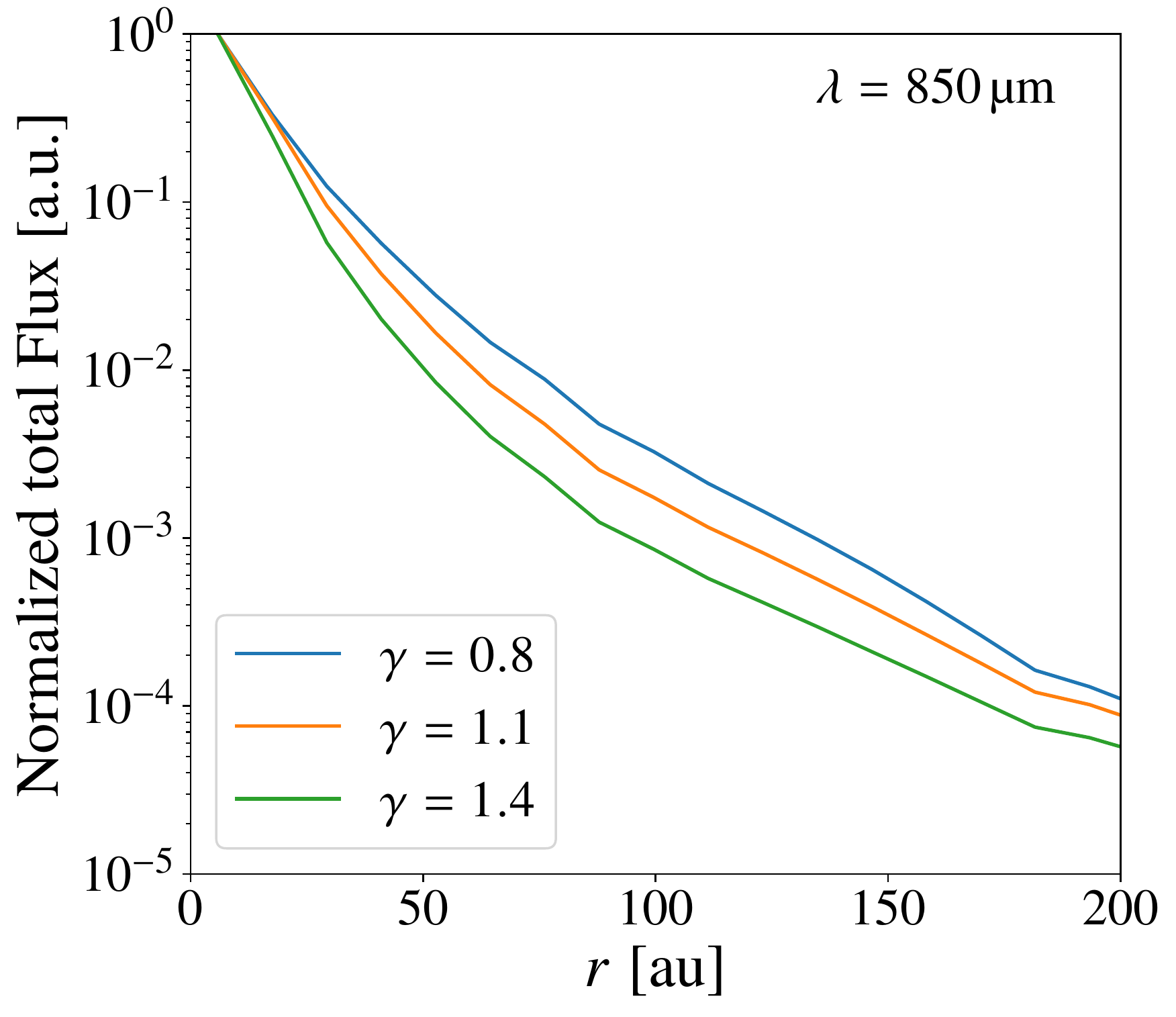}
            \caption{\textit{Left}: Radial profiles of the degree of polarisation only for the scattered radiation of a protoplanetary disk for different values of the radial density exponent $\gamma$. \textit{Right}: Normalised intensity profile of re-emitted plus scattered flux. A higher value of the exponent $\gamma$ corresponds to a steeper density distribution, and the intensity gradient is larger. The gradient of the intensity profile may serve as a qualitative measure for the anisotropy of the radiation field. Larger gradients lead to a higher degree of anisotropy and thus to a higher degree of polarisation.}
            \label{fig:P_sca}
        \end{figure}

        In \citet{kataoka-et-al-2014}, the authors show that the polarisation degree of self-scattering is given by the anisotropy of the radiation field at the point of scattering. Put simply, for a fixed ratio of scattered to direct re-emission radiation, the polarisation degree is maximised if the radiation field only has components  along one axis, and is minimised if the radiation field is isotropic. In the case of a protoplanetary disk, the radiation field has mostly radial components pointing away from the inner disk regions because the inner part of the disk is usually brighter than the outer parts. Thus, the radiation field becomes less isotropic for larger radii and the polarisation degree of purely scattered light (i.e. no direct re-emitted flux) increases. This can be seen in the left-hand panel of \pref{fig:P_sca} and it was also mentioned for the mid-infrared by \citet{heese-wolf-brauer-2020}. For steeper density distributions (i.e. larger values of the exponent $\gamma),$ the anisotropy and, thus, the polarisation degree increase when the flux ratio of the inner to the outer part of the disk is larger (compared to a smaller exponent). As a rough measure for the radiation field, we show the normalised total flux over distance $r$. For a wavelength of $\lambda=\SI{850}{\um}$, the radial profile of the flux can be seen in the right-hand panel of \pref{fig:P_sca}. A more shallow density distribution shows a smaller gradient and the radiation field in the disk is more isotropic.

    \subsection{Optical depth}
    \label{sec:res_tau}
        The influence of the optical depth on the appearance of a protoplanetary disk, as it concerns the scope of this study, is threefold.

        \subsubsection{Scattering probability}
\begin{figure}
                \includegraphics[width=0.495\linewidth]{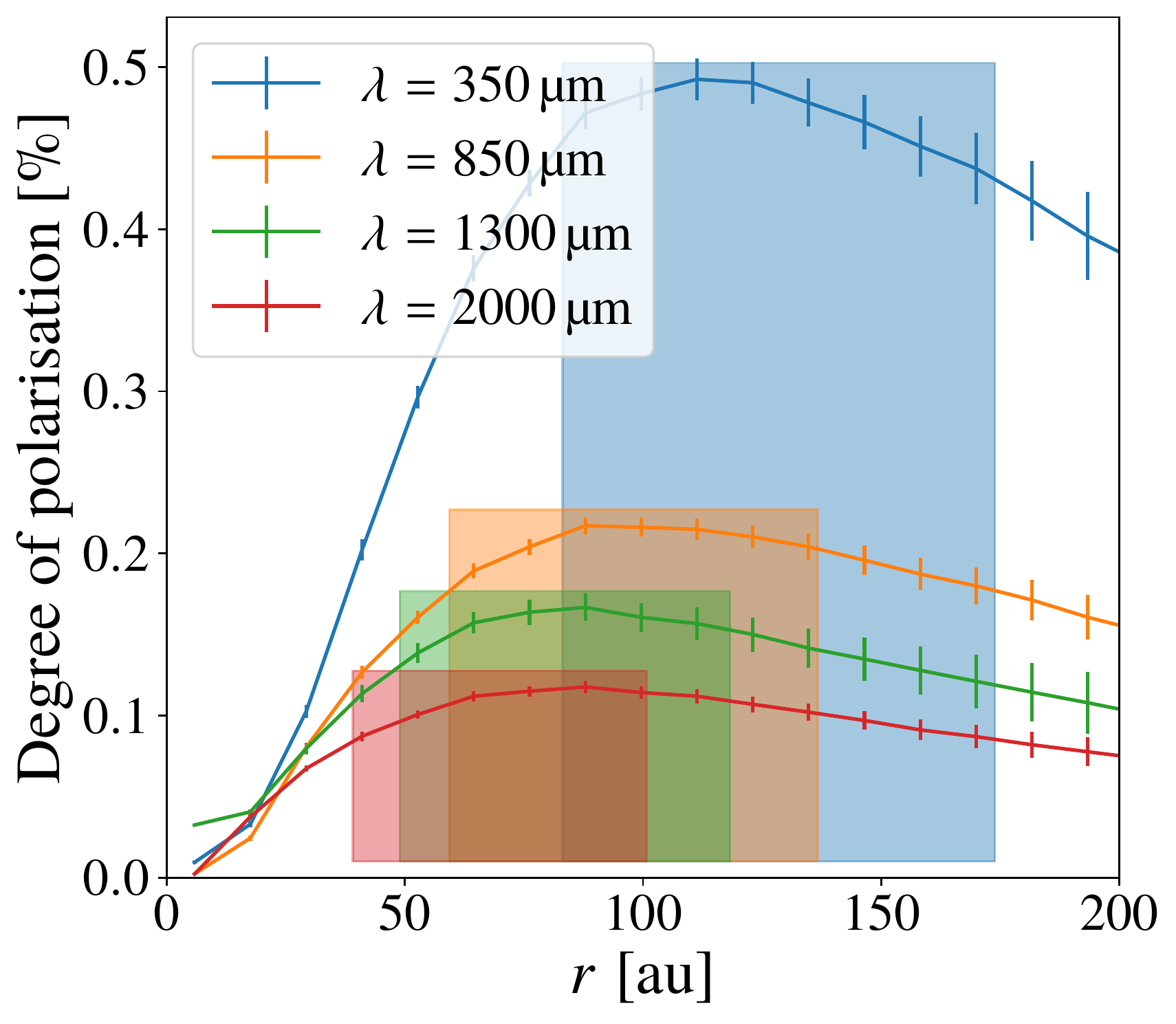}
                \includegraphics[width=0.495\linewidth]{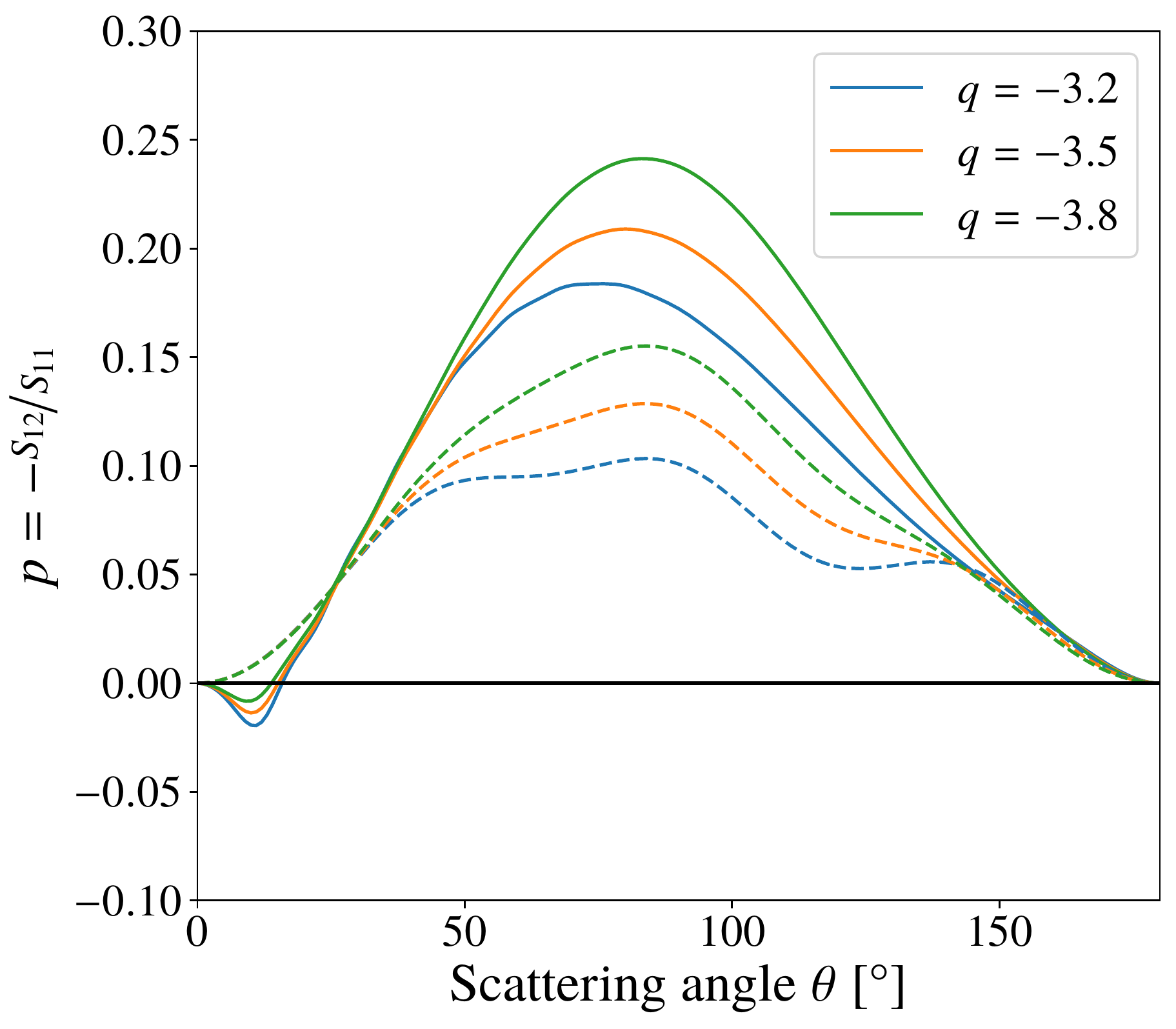}
                \caption{\textit{Left}: Radial profiles of the degree of polarisation due to scattering of a protoplanetary disk seen face-on for different wavelengths $\lambda$ (reference model). The intensity includes both direct and scattered thermal radiation. The shaded areas show the radial distance where the optical depth through the disk to the observer is in the range between \num{0.1} and \num{0.02}. \textit{Right}: Analytical polarisation degree of a single scattering event $p = -\nicefrac{S_{12}}{S_{11}}$ for the wavelengths $\lambda=\SI{350}{\um}$ (\textit{solid lines}) and $\lambda=\SI{2}{\mm}$ (\textit{dashed lines}), averaged over the entire grain size range from \SI{5}{\nm} to \SI{1}{\mm} for different values of the exponent $q$ of the grain size distribution.}
                \label{fig:P_wave_theta}
            \end{figure}

        In general, the probability for scattering along the path of a photon is given by $1-\exp{\left(-\tau_{\text{sca}}\right)}$, with the scattering optical depth $\tau_{\text{sca}}$ between the points of emission and scattering. In regions of high optical depth ($\tau_{\text{sca}}\gg1$), the scattering probability is close to unity. The polarisation degree due to scattering in these regions is thus a function of the albedo (see \pref{sec:res_amax}) and, for anisotropic scatterers, also a function of the scattering angle. In regions with a very low optical depth, $\tau_{\text{sca}}\ll1$, the scattering probability is roughly equivalent to the optical depth, $1-\exp{\left(-\tau_{\text{sca}}\right)} = 1 - \left(1 - \tau_{\text{sca}} + \mathcal{O}(\tau_{\text{sca}}^2)\right) \approx \tau_{\text{sca}}$. Therefore, the ratio of scattered radiation, which is the product of scattering probability, albedo, and flux of the initial re-emitted radiation, to direct radiation is lower, and the polarisation degree is lower as well. As a consequence, the ratio of scattered to direct flux decreases for larger distances from the star as the optical depth decreases. At the same time, the polarisation degree is maximised at radii that roughly coincide with the transition from optically thick to optically thin if the albedo is high. The flux ratio varies between different disks because the vertical density and intensity structure depend on the actual disk set-up. As a consequence, only a range of the optical depth, rather than a fixed value, can be given; thus, the location where the degree of polarisation is maximised is at slightly different radii $r$. In addition to the optical depth $\tau_{\text{sca}}$ between the points of emission and scattering, the (extinction) optical depth through the disk towards the observer $\tau_{\text{obs}}$ also decreases the observed flux. This decrease is identical for both the direct re-emission and the scattered radiation. These optical depths are proportional to each other, and the optical depth towards the observer can be used as a proxy for the scattering probability in the disk. In the left-hand panel of \pref{fig:P_wave_theta}, the shaded areas represent the region where $0.02 \leq \tau_{\text{obs}} \leq 0.1$, which coincides with the maximum degree of polarisation.

\begin{figure}
                \includegraphics[width=0.495\linewidth]{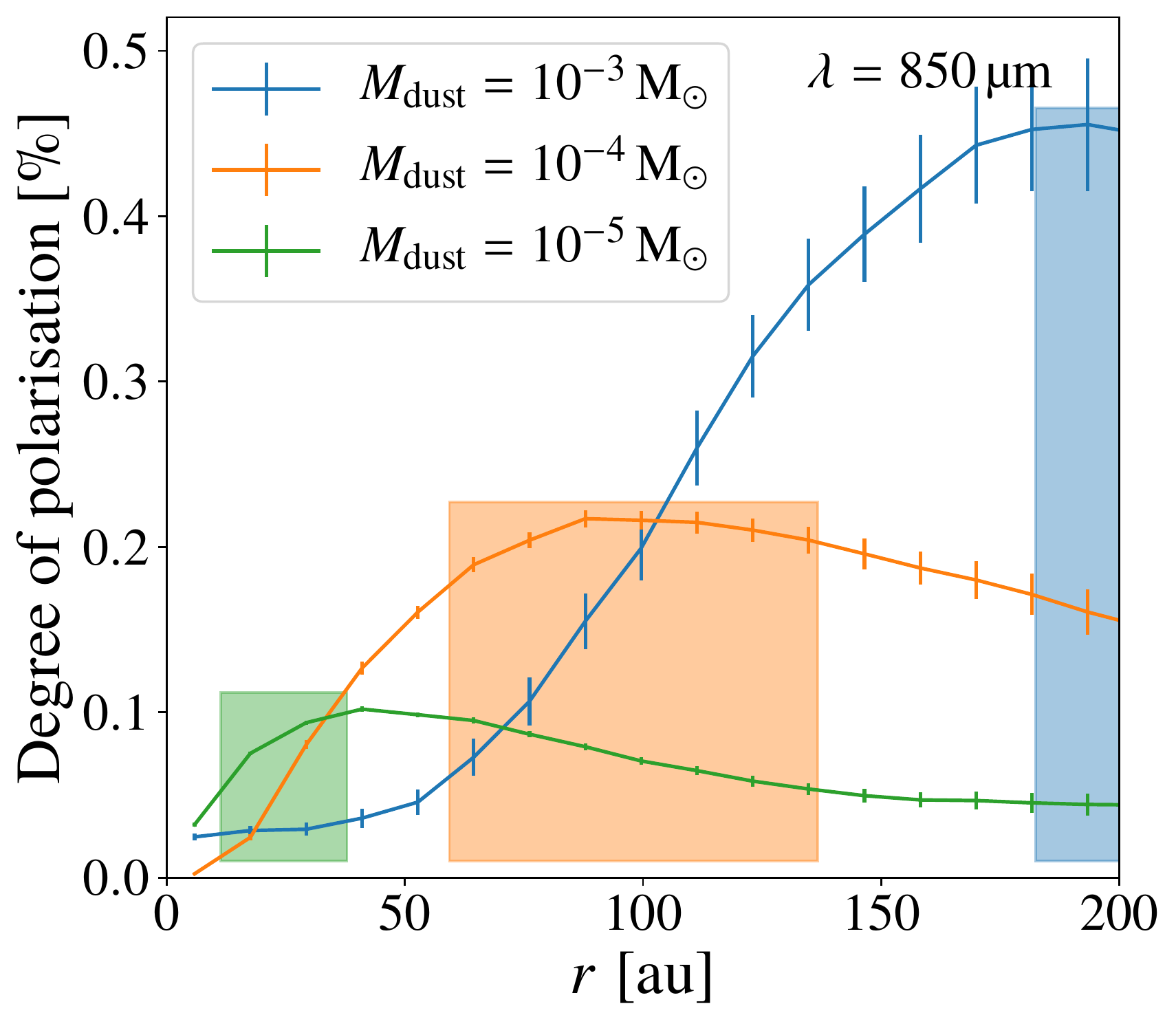}
                \includegraphics[width=0.495\linewidth]{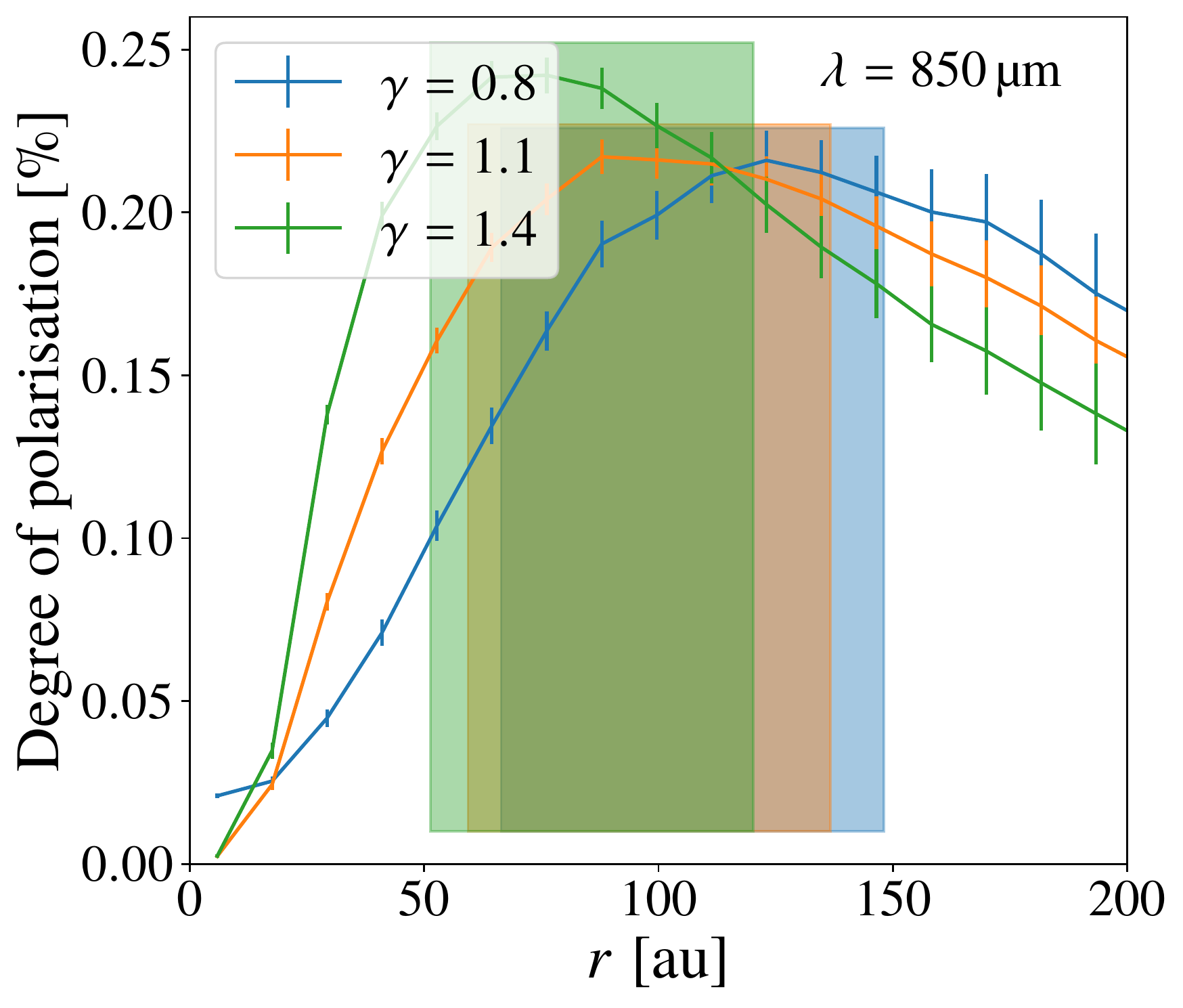}
                \caption{Radial profile of the degree of polarisation for different values of the dust mass (\textit{left}) and radial density exponent $\gamma$ (\textit{right}). The shaded areas coincide with the radial distance of the transition from optically thick to optically thin. For details, see \pref{fig:P_wave_theta}.}
                \label{fig:mass_rad_expo}
            \end{figure}

            The simplest way to increase the optical depth is by increasing the total dust mass of a disk. The peak of the polarisation degree is shifted to larger distances because the transition region from optically thick to optically thin is also shifted to larger distances (see the left-hand panel of \pref{fig:mass_rad_expo}).

            Furthermore, if the exponent $\gamma$ of the radial distribution of the surface density is increased, more mass is shifted closer to the star. This results in a higher optical depth at shorter distances and a lower optical depth (per geometrical path length) in the outer regions of the disk. In the right-hand panel of \pref{fig:mass_rad_expo}, the radial profile of the polarisation degree is shown for different values of the exponent $\gamma$. Since the polarisation degree peaks roughly at the point of transition between optically thick and optically thin, this peak is closer to the star for larger values of the exponent $\gamma$. This effect is most prominent for longer wavelengths. Furthermore, for steeper radial density distributions (i.e. larger values of the exponent $\gamma$), the anisotropy of the radiation field increases and thus so does the degree of polarisation (see \pref{sec:res_radfield}).

        \subsubsection{Observable regions}
            For regions with high optical depths, most of the flux that reaches the detector emerges from regions where $\tau_{\text{obs}}\lesssim1$. These upper layers of the disk are hotter than the heavily shielded midplane, and the local grain size distribution is different for different vertical layers due to the settling of the largest grains. A change in the total optical depth, for example from a change in the total disk mass, shifts the $\tau_{\text{obs}}=1$-plane to different vertical regions in the disk. Due to the dust settling, this leads to different local grain size distributions, and different grain sizes with different optical properties contribute to the total flux. Thus, the overall appearance of the disk may change drastically.

        \subsubsection{Stellar heating}
            Lastly, the optical depth measured from the star towards the disk $\tau_{\star}$ determines the temperature of the dust. For very high optical depths $\tau_{\star}$ (e.g. in the midplane of the inner few $\si{au}$ of the disk), the temperature is mainly a product of 'back-warming' from the directly heated, adjacent dust grains, which radiate at much longer wavelengths compared to the star. However, at the inner rim and upper layers of the disk, where the optical depth is small, the temperature is substantially higher. These areas, which emit the highest amount of energy per unit dust mass, will be at different places in the disk and have different extents if the optical depth $\tau_{\star}$ is changed.

            Different flarings of a disk cause very different temperature structures. Stronger flaring leads to a smaller scale height in the inner regions but to larger scale heights in the outer regions of the disk. The vertical geometric cross-section of the disk is smaller close to the star, and therefore the optical depth is larger and the temperatures are lower compared to the reference case. Further away from the star, the vertical cross-section of the disk increases as flaring increases, as do temperatures with respect to the reference model. Overall, the temperature distribution is more shallow for more strongly flared disks, and the radiation field is more isotropic, leading to less polarisation. The radial profiles of the polarisation degree for different values of the flaring parameter $\beta$ can be found in the upper-left panel of \pref{fig:size_dist}.

            For larger scale heights \href{}, the stellar heating is more efficient because the optical depth as seen from the star is decreased. This enhances the dust temperature, and the anisotropy of the radiation field increases (see \pref{sec:res_radfield}), as does the degree of polarisation; see the middle-left panel of \pref{fig:size_dist}.

            The viscosity of the disk $\alpha$ controls the magnitude of the dust settling and thus the scale height of the larger grains. For highly viscous disks, characterised by large values of $\alpha$, the scale height of the large dust grains is increased (see \pref{eq:settling}). Therefore, the polarisation degree is largest if the settling is weakest (i.e. for the most viscous disks); see the bottom-left panel of \pref{fig:size_dist}.

    \subsection{Observing wavelength}
    \label{sec:res_wave}
        In the left-hand panel of \pref{fig:P_wave_theta}, the radial profile of the polarisation degree shows a decrease in the polarisation degree with increasing wavelength $\lambda$. This decrease is independent of the actual disk set-up, and thus we only present the results for the reference model. Increasing the wavelength causes a series of effects that determine the net polarisation degree.

        First, the optical depth decreases due to smaller extinction efficiencies. The effect of a smaller optical depth is described in \pref{sec:res_tau}.

        Second, for longer wavelengths, dust with lower temperatures contributes more to the overall flux, as the amount of cold dust is higher than for hot and warm dust. Thus, the flux ratio of the outer to inner disk becomes more balanced and this causes a smaller anisotropy of the radiation field, which is addressed in \pref{sec:res_radfield}.

\begin{figure}
            \includegraphics[width=0.495\linewidth]{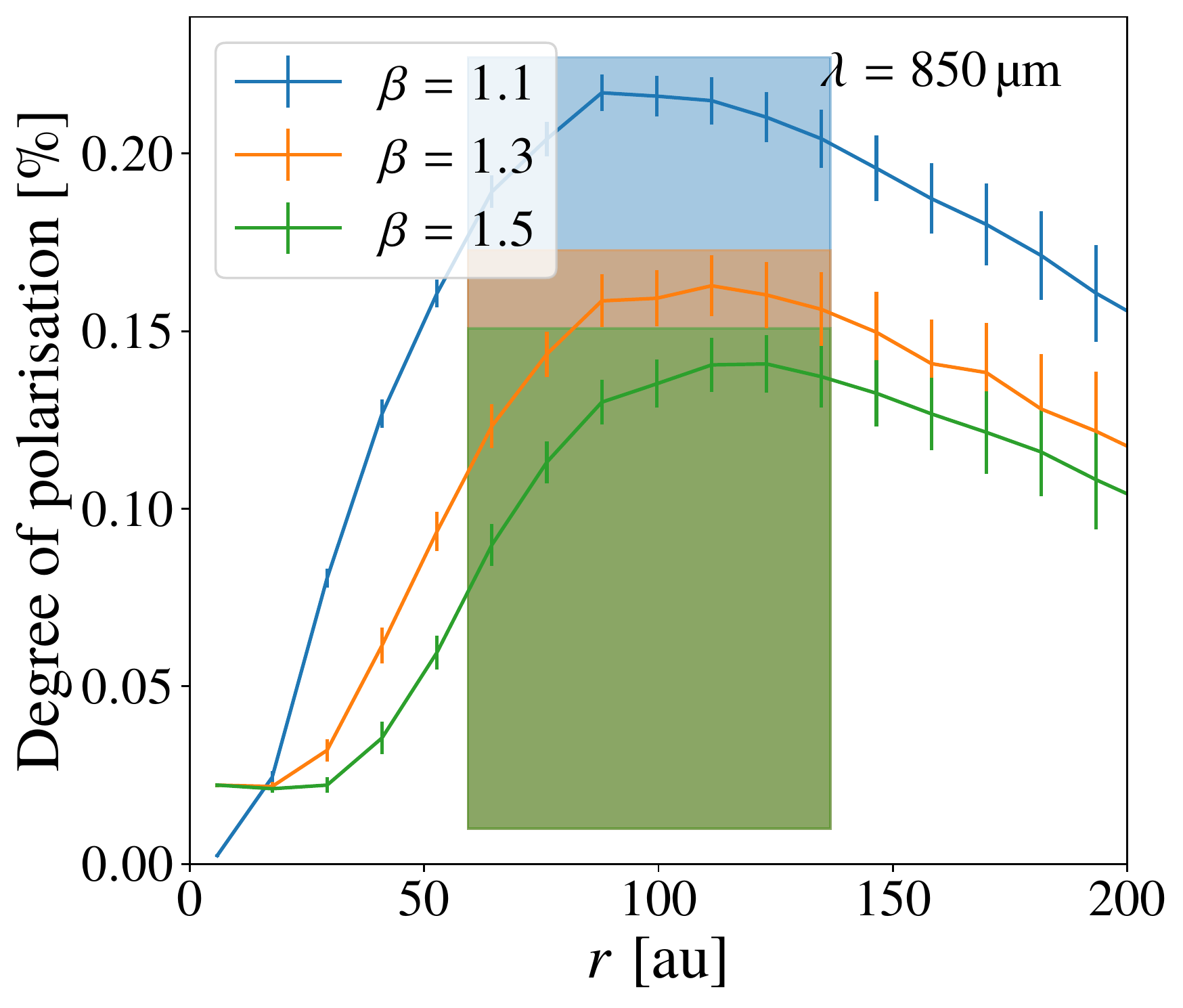}
            \includegraphics[width=0.495\linewidth]{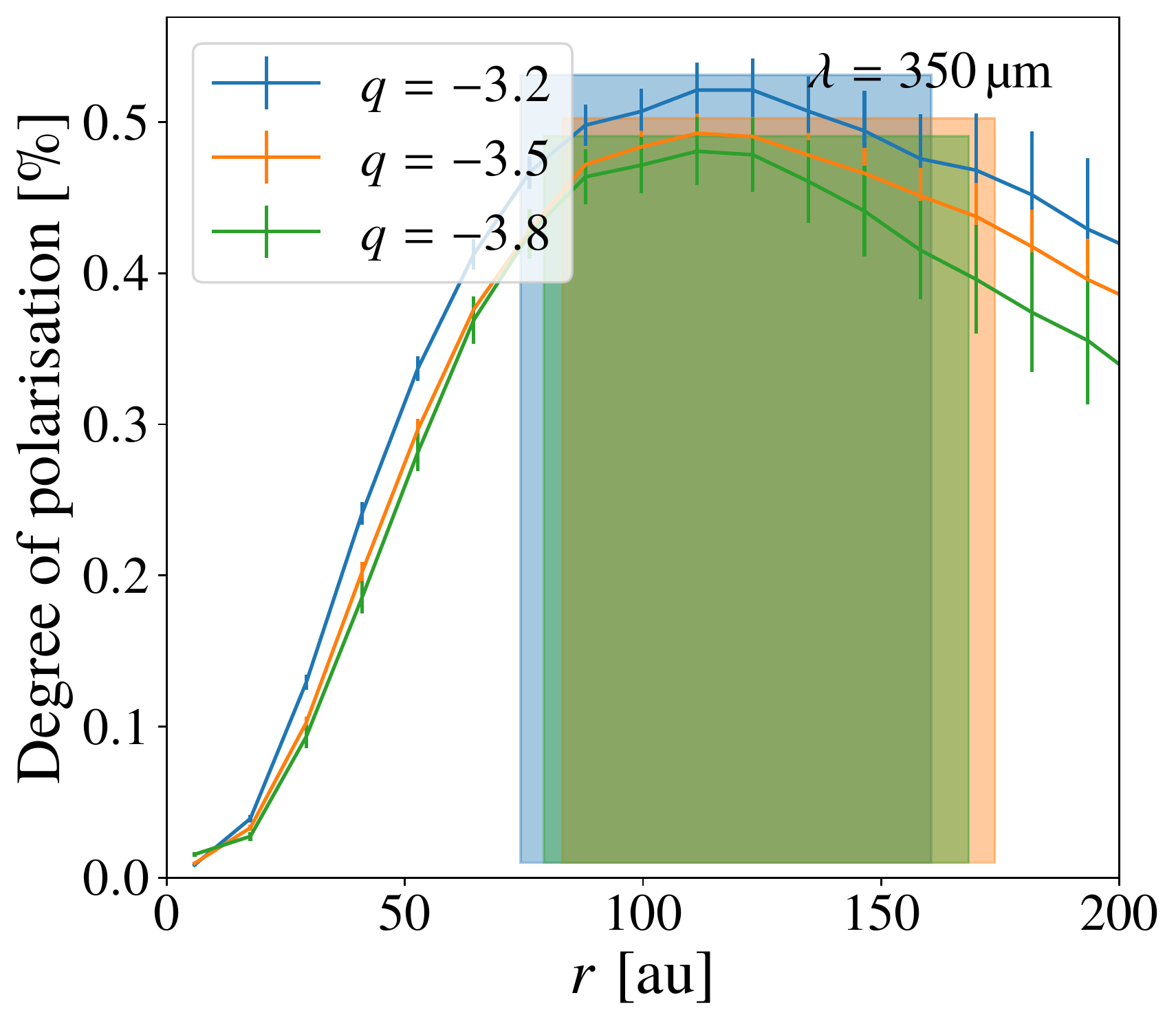}\\
            \includegraphics[width=0.495\linewidth]{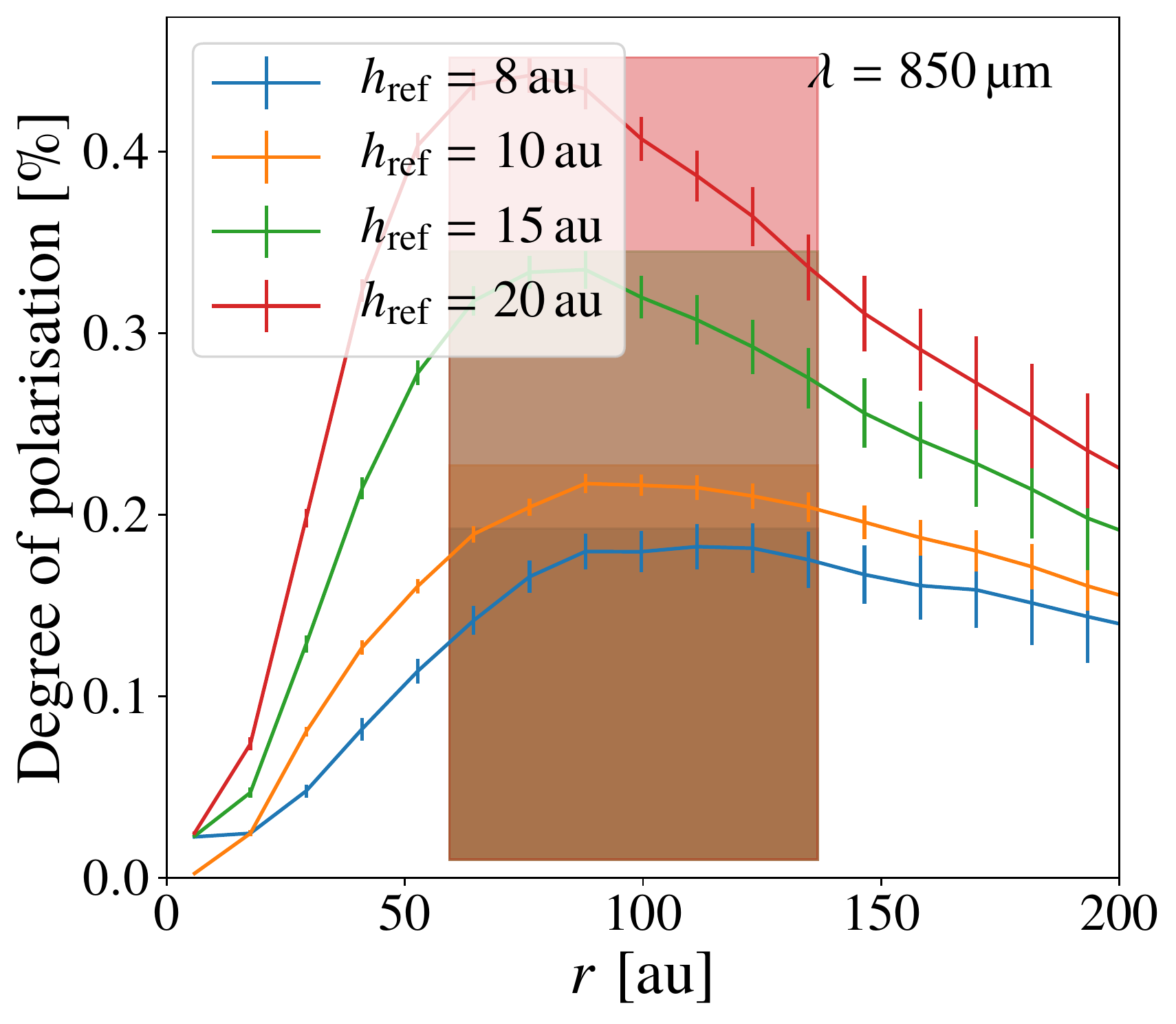}
            \includegraphics[width=0.495\linewidth]{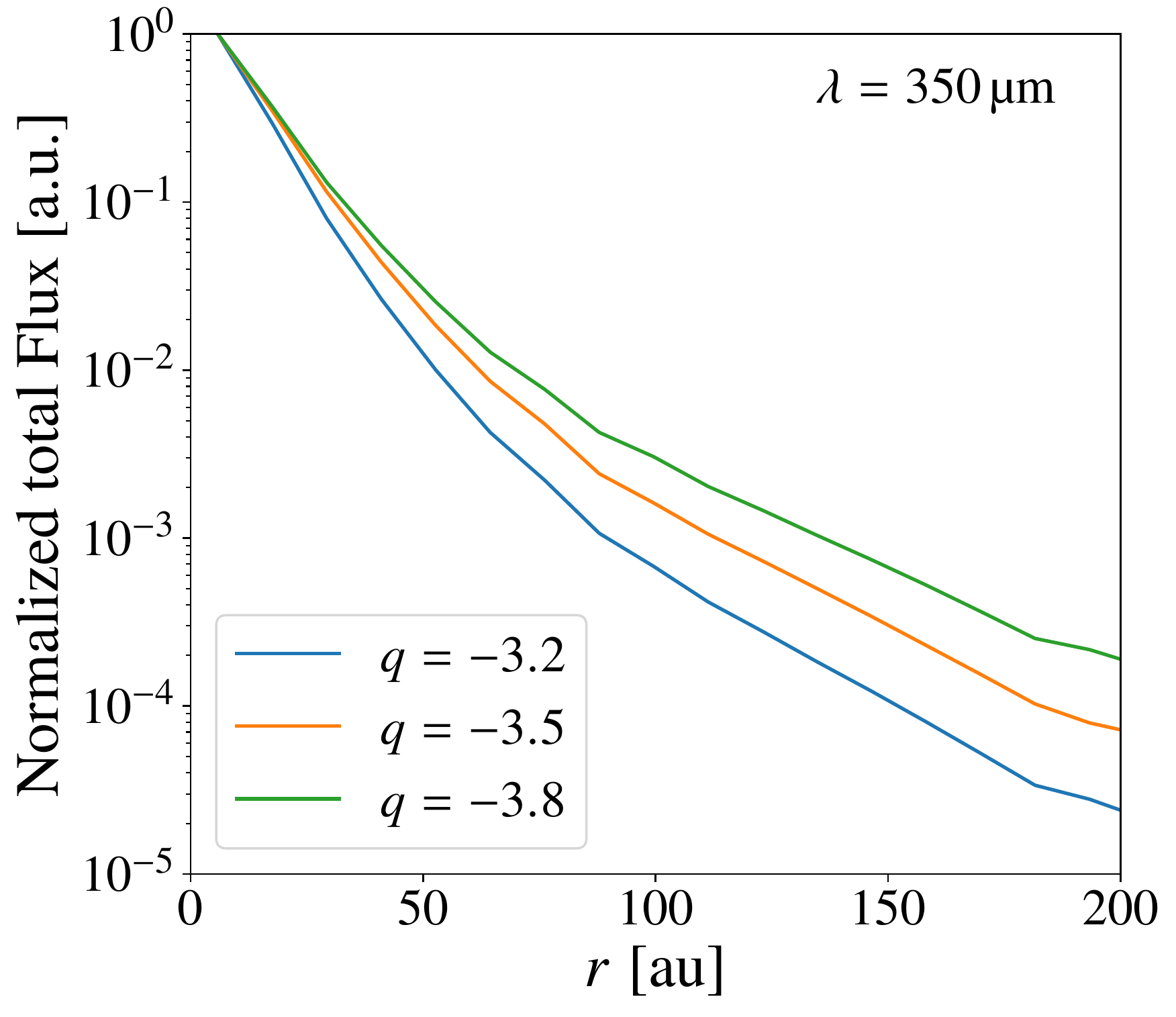}\\
            \includegraphics[width=0.495\linewidth]{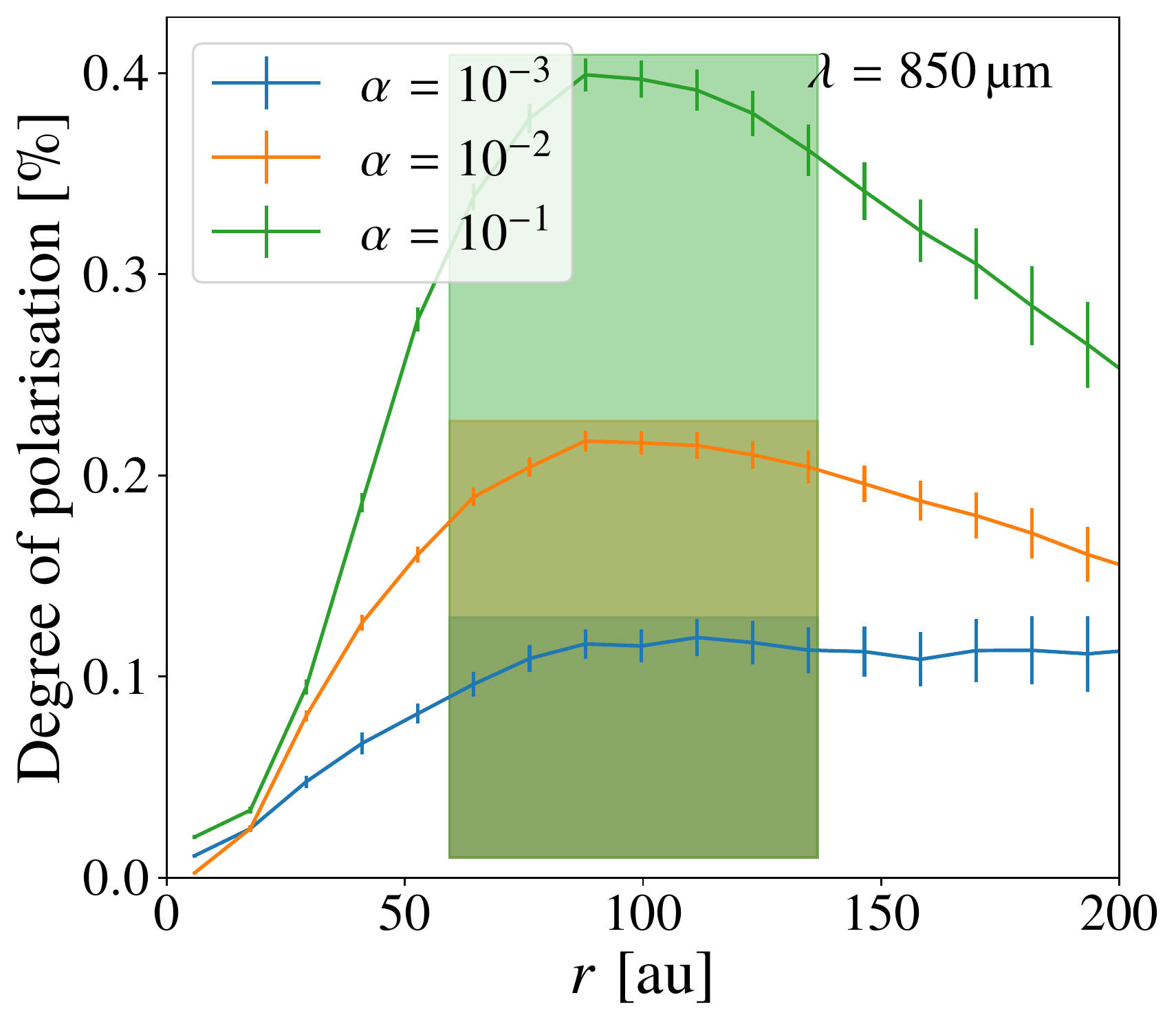}
            \includegraphics[width=0.495\linewidth]{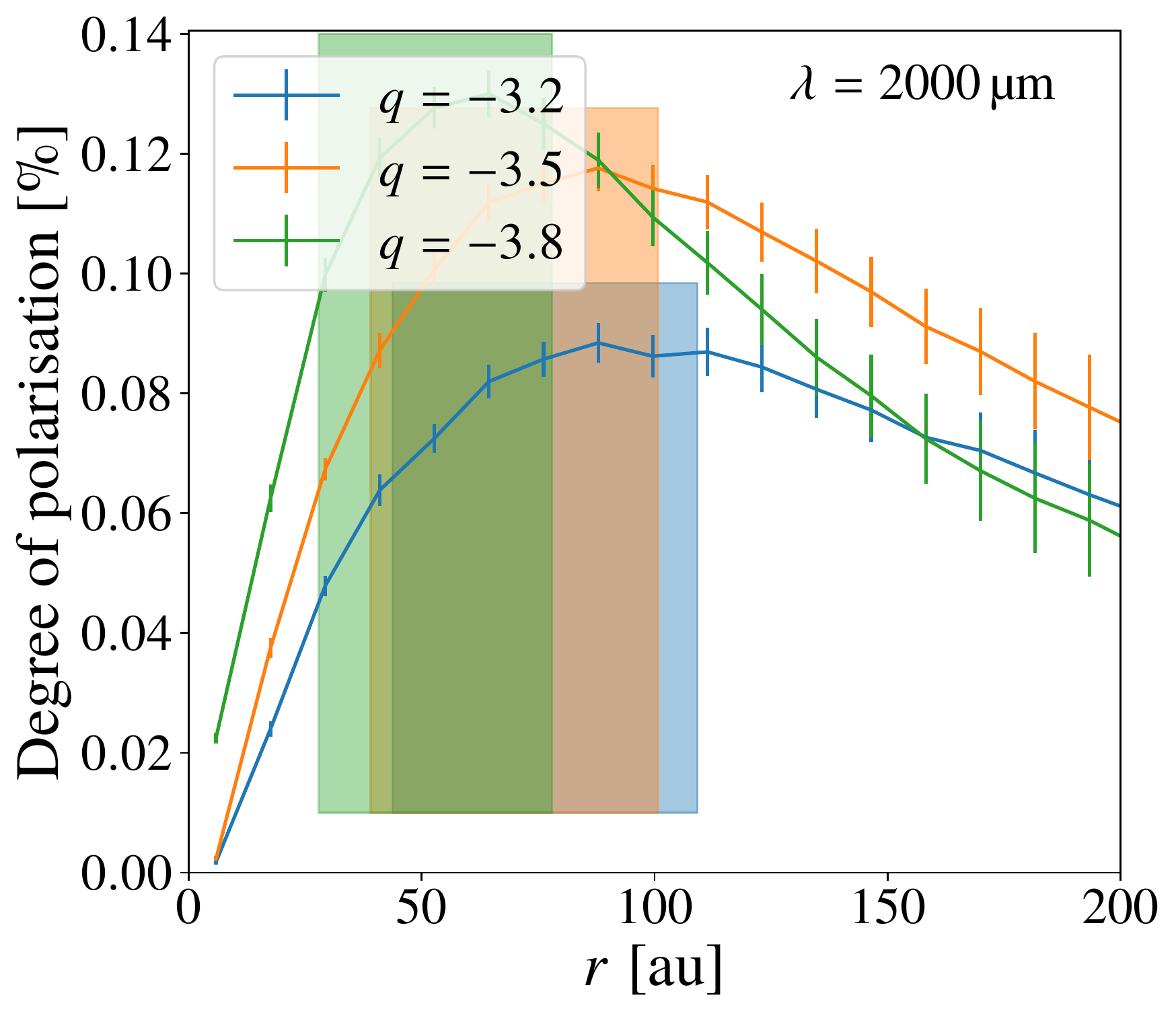}
            \caption{Radial profile of the degree of polarisation for different values of the flaring parameter $\beta$ (\textit{upper left}), the scale height \href{} (\textit{middle left}), the viscosity parameter $\alpha$ (\textit{lower left}), and the exponent of the grain size distributions (\textit{upper} and \textit{lower right}). The shaded areas coincide with the radial distance of the transition from optically thick to optically thin. For details, see \pref{fig:P_wave_theta}. \textit{Middle right}: radial profile of the normalised total flux for different grain size distributions at $\lambda=\SI{350}{\um}$.}
            \label{fig:size_dist}
        \end{figure}

        Third, the optical properties of the dust mixture change. The albedo and the flux ratio of scattering to direct re-emission increase. At the same time, the transition from Rayleigh to Mie scattering (i.e. where $s\approx\lambda$) happens for larger grain radii $s$. This leads to a decrease in the polarisation degree $p = -\nicefrac{S_{12}}{S_{11}}$ for single scattering (see the right-hand panel of \pref{fig:P_wave_theta}). Since the polarisation degree of the total flux depends on the product of the albedo and $p$, it has a maximum at $s \approx \frac{\lambda}{2\pi}$ \citep[see also][]{kataoka-et-al-2015}. The individual contributions of each grain size must be integrated according to the grain size distribution to investigate the entire disk. We considered an inverse power law for the grain size distribution, $n\propto s^{-q}$ (see \pref{sec:set-up}), and thus smaller grains contribute much more to the total number of dust grains than larger ones do. For smaller wavelengths, the fraction of dust grains with $s \approx \frac{\lambda}{2\pi}$ is larger than in the case of longer wavelengths. This results in a lower overall degree of polarisation for longer wavelengths; see \pref{fig:P_wave_theta}. For the same reason, the polarisation degree of a disk increases (decreases) if the grain size distribution is steeper (more shallow), in other words, if the absolute value of the grain size exponent $q$ is larger (smaller). This can be seen in the right-hand panel of \pref{fig:P_wave_theta} and in the lower-right panel of \pref{fig:size_dist}.

        The upper-left panel of \pref{fig:size_dist} shows that the polarisation degree at a wavelength of $\lambda=\SI{350}{\um}$ is highest for an exponent $q = -\num{3.2}$, which seems to contradict the previous line of reasoning. The optical properties are such that the polarisation degree should be higher for a steeper size distribution. However, for a flat distribution, the absolute number of large grains is higher for a constant total disk mass. Consequently, the volume density $\rho$, and hence the optical depth per geometrical path length, strongly increases in the midplane and decreases in the upper layers of the disk because large grains have a lower scale height. Due to the reduced optical depth, the dust temperature is higher for the smallest grains if compared to the case of a steeper grain size distribution. Therefore, the radial brightness profile of the total emission at shorter wavelengths is much steeper; see the middle-left panel of \pref{fig:size_dist}. Both lead to a more pronounced anisotropy of the radiation field, resulting in this somewhat counterintuitive behaviour for the polarisation degree at $\lambda=\SI{350}{\um}$.

    \subsection{Maximum grain size}
    \label{sec:res_amax}
\begin{figure}
            \includegraphics[width=0.495\linewidth]{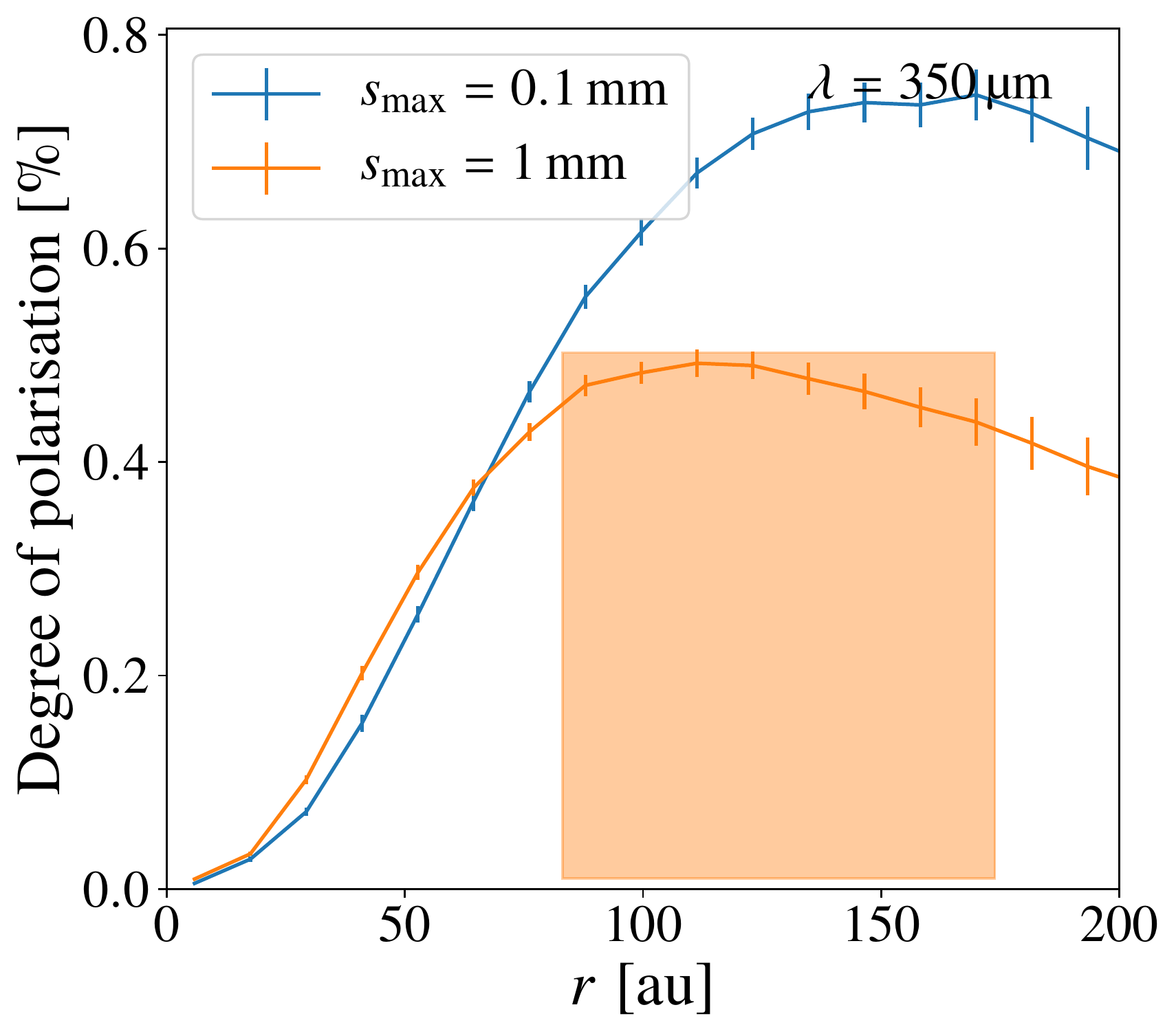}
            \includegraphics[width=0.495\linewidth]{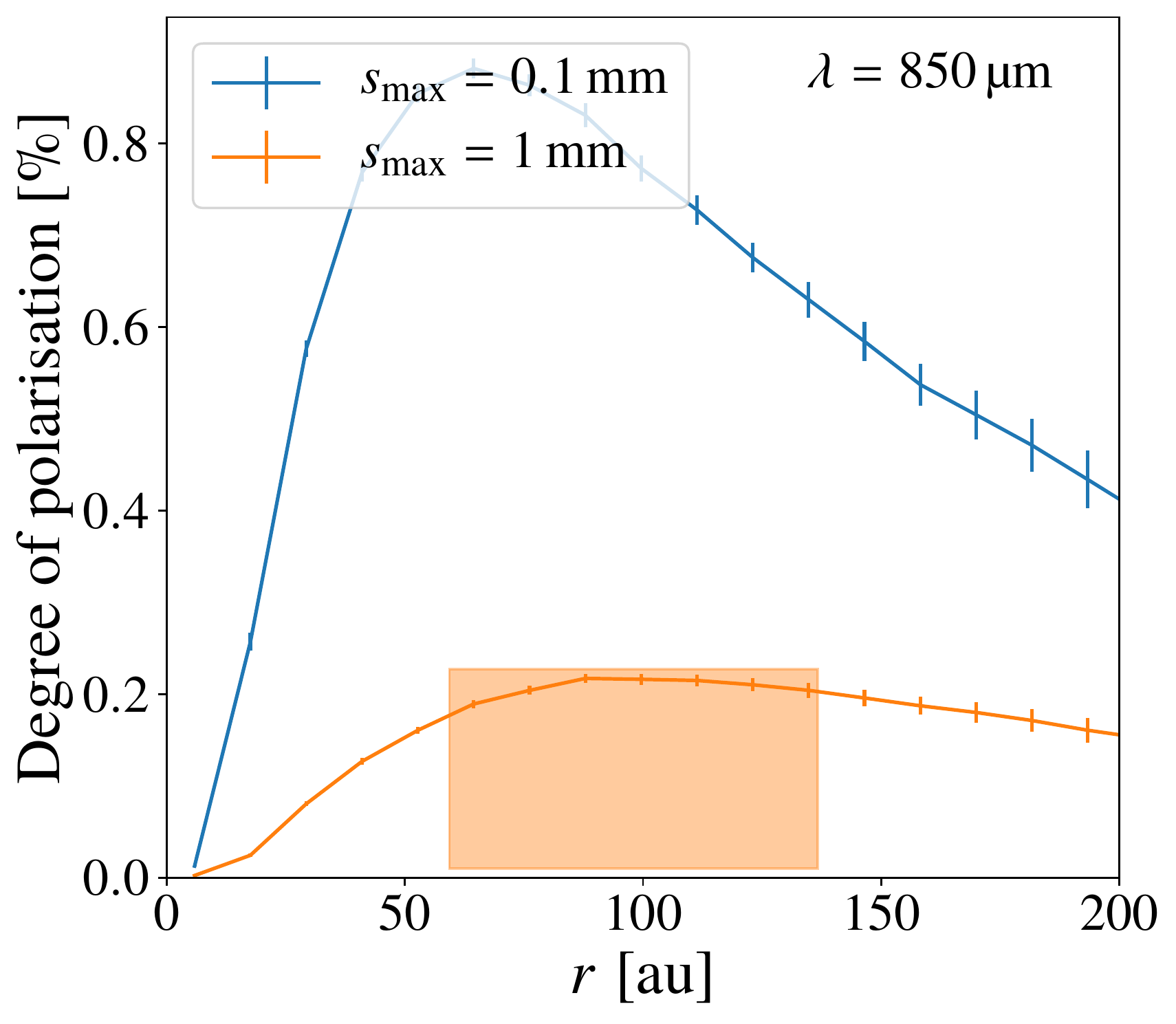}\\
            \includegraphics[width=0.495\linewidth]{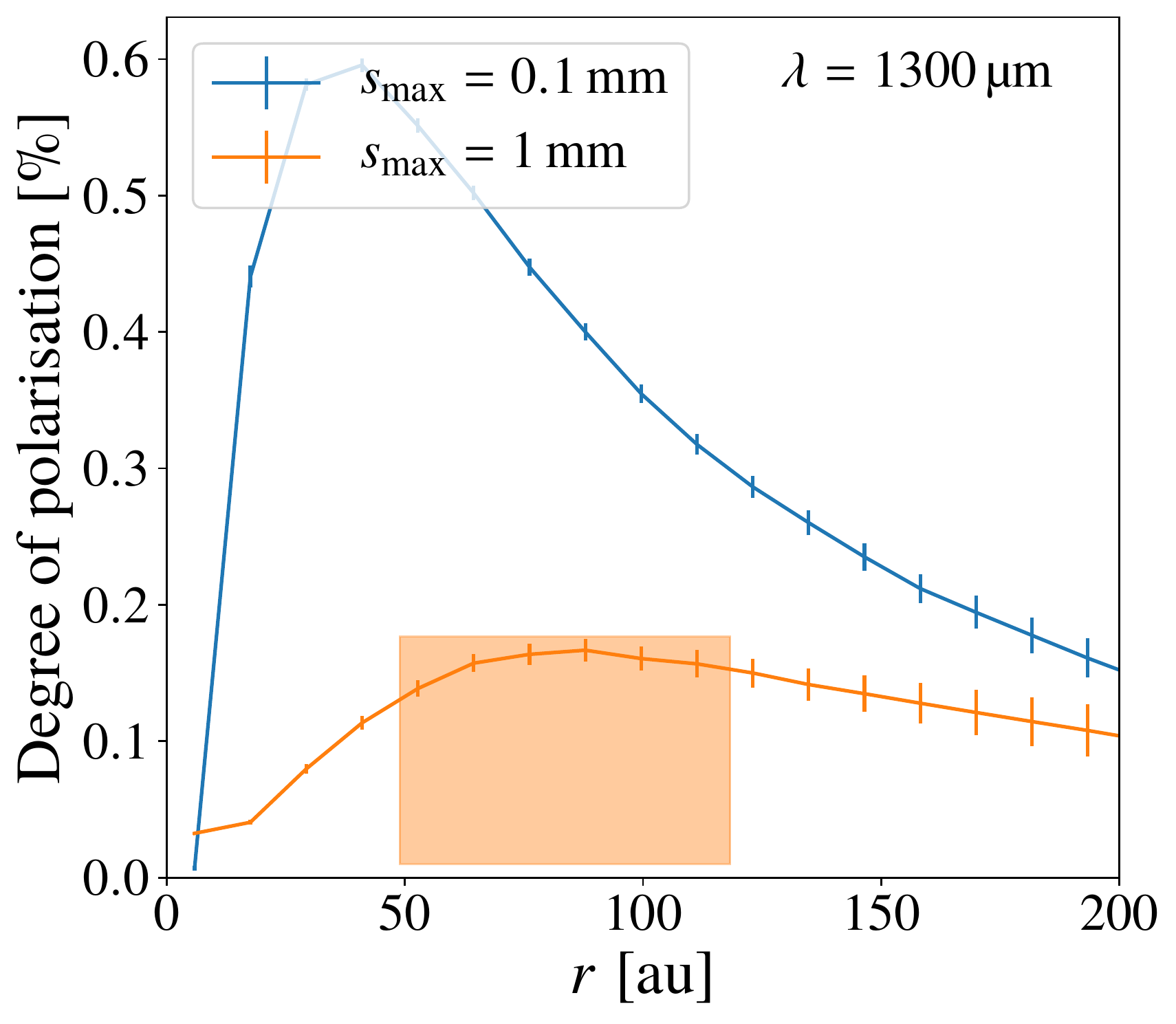}
            \includegraphics[width=0.495\linewidth]{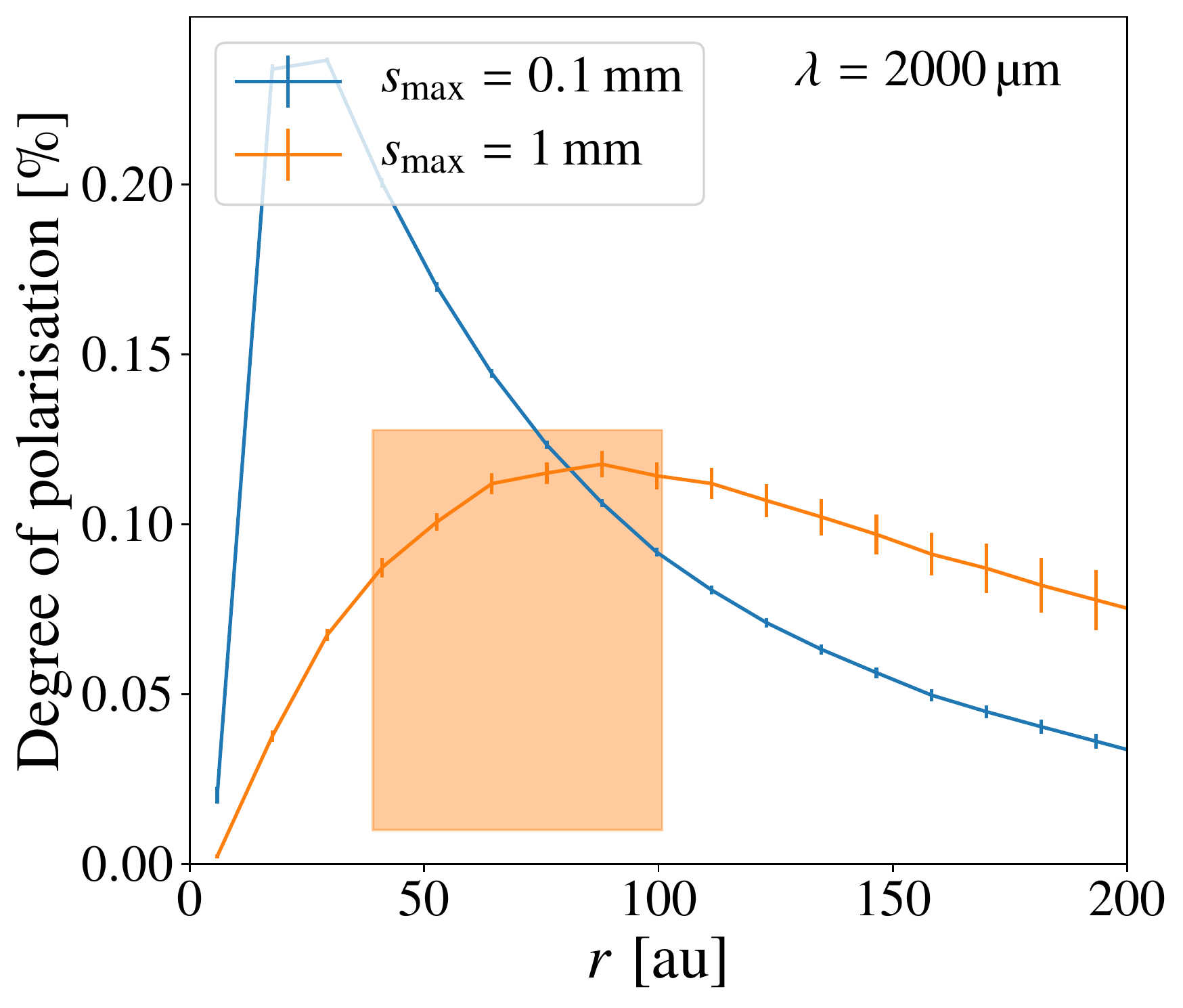}
            \caption{Radial profile of the degree of polarisation for different upper grain size limits. The shaded areas coincide with the radial distance of the transition from optically thick to optically thin. For details, see  \pref{fig:P_wave_theta}. For $a_{\text{max}}=\SI{0.1}{\mm}$, the minimum optical depth is about \num{0.9} at the wavelength $\lambda=\SI{2}{\mm}$.}
            \label{fig:amax}
        \end{figure}

        As mentioned above, the degree of polarisation is maximised for grain sizes roughly of the order of $\frac{\lambda}{2\pi}$ \citep{kataoka-et-al-2015}. Therefore, the grain size distribution with a maximum grain size of $a_{\text{max}}=\SI{0.1}{\mm}$ produces a higher (peak) polarisation degree for all considered wavelengths (see \pref{fig:amax}). The flux ratio of direct to scattered re-emission increases with increasing distance to the star, and the polarisation degree is reduced. This effect is stronger for larger wavelengths because the scattering efficiency decreases after a peak value where $s\approx\frac{\lambda}{2\pi}$, whereas the emission efficiency increases. This leads to low values for the size-integrated albedo; for example, the albedo for a wavelength of \SI{2}{\mm} is: \num{0.83} for $a_{\text{max}}=\SI{1}{\mm}$; and 0.04 for $a_{\text{max}}=\SI{0.1}{\mm}$. The distance $r$ of the maximum polarisation degree is therefore not at the transition from optically thick to optically thin, because only a very small amount of the emitted light is scattered.

        A (maximum) grain size of $\sim\SI{100}{\um}$ is often discussed when interpreting polarisation observations in the sub-mm regime \citep{ohashi-kataoka-2019,dent-et-al-2019,hull-et-al-2018,kataoka-et-al-2017,yang-et-al-2016a}. This is because the polarisation efficiency is maximised for this wavelength-grain size combination, as explained above. This contradicts the usual finding of very large grains if deduced from the spectral index at mm wavelengths. Proposed solutions to this contradiction include grain porosity and different compositions (see also \citealt{yang-li-2020,kataoka-et-al-2017,kataoka-et-al-2015}). Keeping the maximum grain size of \SI{1}{\mm}, we calculated the grain-size averaged degree of polarisation of a single scattering event $p = -\,\nicefrac{S_{12}}{S_{11}}$ for different mass fractions of silicate in the dust composition, and for different grain porosities. Our reference grain model is composed of compact ($P=0$) grains made up of \SI{62.5}{\percent} silicate and \SI{37.5}{\percent} graphite (see \pref{sec:set-up}). An increase in both the graphite mass fraction and the grain porosity independently result in an increase in the single-scattering polarisation degree, and thus in an increase in the degree of polarisation of the entire disk, including mm-sized grains. Additionally, the magnitude of this increase is wavelength-dependent and may help avoid the significant drop in the polarisation degree with increasing wavelengths, as can be seen, for instance, in \pref{fig:P_wave_theta} (left).
        \begin{figure}
            \includegraphics[width=0.495\linewidth,height=0.25\textheight]{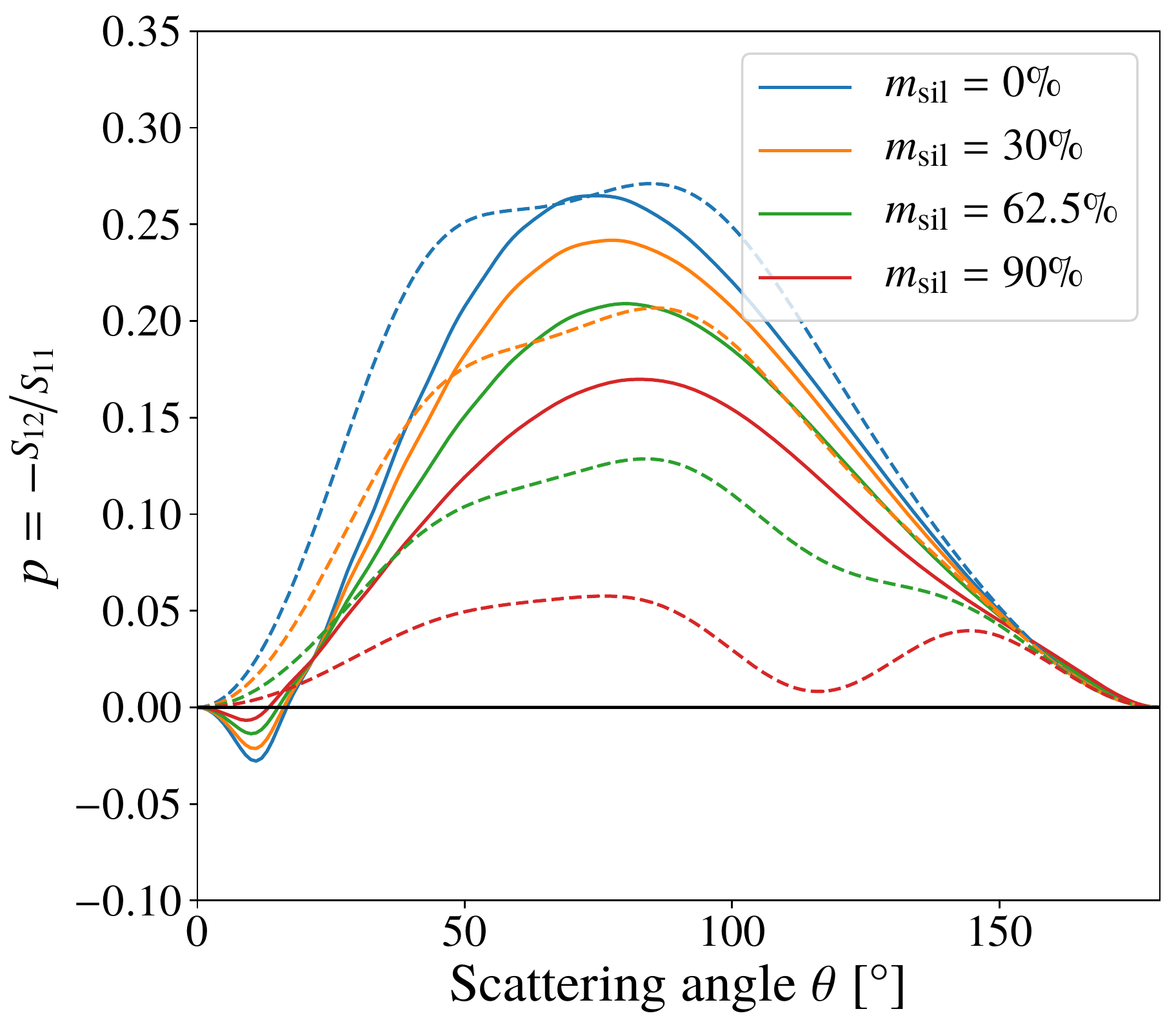}
            \includegraphics[width=0.495\linewidth,height=0.25\textheight]{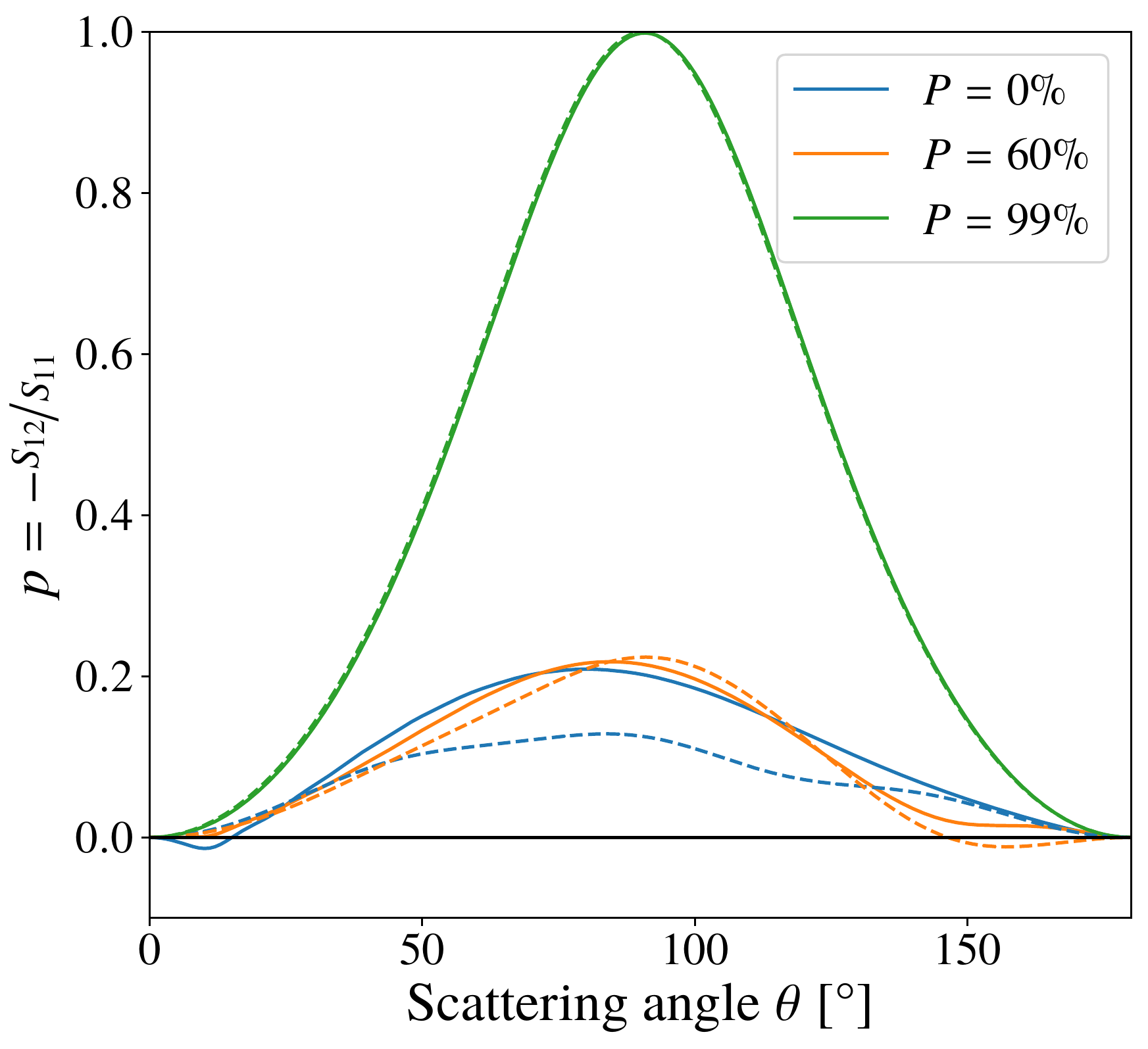}
            \caption{Analytical polarisation degree of a single scattering event $p = -\nicefrac{S_{12}}{S_{11}}$ for different values of silicate mass fraction $m_{\text{sil}}$ (\textit{left}) and porosity $P$ (\textit{right}) for the wavelengths $\lambda=\SI{350}{\um}$ (\textit{solid lines}) and $\lambda=\SI{2}{\mm}$ (\textit{dashed lines}). For details, see \pref{fig:P_wave_theta}.}
            \label{fig:comp_por}
        \end{figure}
\section{Discussion}
\label{sec:discussion}
    \subsection{Constraining disk and dust properties}
        The polarisation degree for scattered re-emitted light from a protoplanetary disk is affected by the various geometrical parameters of the disk as well as the dust properties. Besides the optical depth, the degree of anisotropy of the radiation field is the most vital point, as shown by \cite{kataoka-et-al-2014}. The disk geometry is able to both enhance and reduce this anisotropy with multiple ambiguities. The anisotropy, and hence the polarisation degree, can be increased by altering one or multiple disk parameters; for example, the scale height \href{} and the disk flaring produce very similar results and are thus indistinguishable by the study of polarised light only. The presented results suggest that most of the investigated parameters do not produce specific features in the radial profiles of the polarisation degree, which would have allowed observers to constrain these parameters.

        There is, however, one exception to this conclusion. As can be seen for nearly all radial profiles, the polarisation degree peaks at a certain radius, which is just barely altered for different parameter values. This position of the peak roughly coincides with the radius, where the optical depth through the disk as seen from the observer is reduced to about $\num{0.1}$. The probability of a scattering event in low optical depth regions is roughly equivalent to the albedo multiplied by the optical depth; that is, the scattering probability for grains with high albedos is about \SI{1}{\percent} if the optical depth is \num{0.01}. Consequently, the fraction of the (polarised) scattered light is reduced for low optical depths, and the polarisation degree decreases again for the outer parts of the disk. As the optical depth towards the observer is a function of the surface density $\Sigma_{\text{dust}}$, it is independent of flaring, scale height, and dust settling, and changes only slightly for different values of the grain size exponent $q$. Therefore, the peak position of the polarisation degree is indicative of the dust mass, grain size, and radial exponent $\gamma$. If complementary observational constraints exist, for example the spectral index at mm wavelengths or the spectral energy distribution over a large wavelength range, these parameters may be constrained to a small range of possible values. In particular, the dust mass has a huge impact on the polarisation degree and may be the easiest parameter to derive. The derived mass can then be compared to other, independent mass estimates, for example from the total flux level or the mass absorption coefficient. However, this conclusion can only be drawn if the overall albedo is high enough (i.e. close to one; see \pref{sec:res_amax}). Otherwise, the ratio of direct to scattered re-emission is large even for optically thick regions, and the peak may occur at shorter distances from the star.
    \subsection{Low polarisation degree}
        A very important finding of our parameter study is that the overall degree of polarisation is very low. In fact, it is much lower (by a factor of $\sim$ \num{2} -- \num{10}) than found in recent (sub-)mm disk observations \citep[e.g.][]{ohashi-kataoka-2019,dent-et-al-2019,bacciotti-et-al-2018,ohashi-et-al-2018,hull-et-al-2018, stephens-et-al-2017,stephens-et-al-2014} where the polarisation is thought to come at least partially from scattered re-emission. This is in agreement with the recently published study by \citet{yang-li-2020}; in their model, only pure carbonaceous material can produce a high enough polarisation degree in a size distribution with a maximum grain size of \SI{3}{\mm} to explain at least some of the observations. We find similar results when decreasing the fractional amount of silicate or increasing the grain porosity. Another possible explanation might be a more complex grain shape. In a recent study, \citet{kirchschlager-bertrang-2020} focused on oblate silicate grains and find that in some cases the single scattering polarisation is substantially higher than for spherical grains.

        Our reference disk model used here is based on extensive results from both theoretical and observational works concerning protoplanetary disks and dust evolution carried out over several decades. Although it is possible to maximise the polarisation degree in our simulations to get closer to the values derived by observations by using a combination of extreme values within our parameter space (e.g. large scale height, weak flaring, or high viscosity), it is highly improbable that all observed disks show such extreme density distributions. Therefore, the effect of different underlying models of disk density distributions, dust settling, and dust compositions and shapes on the resulting polarisation of protoplanetary disks needs to be addressed in future studies.

\section{Summary}
\label{sec:summary}
    We investigated the polarisation due to scattered thermal re-emission in protoplanetary disks where large dust grains have already settled closer to the midplane. The amount and spatial distribution of polarised light were calculated using the versatile, publicly available Monte Carlo radiative transfer code \texttt{POLARIS}.

    With the adapted model for the density distribution, dust composition and settling, and disk geometry parameters, we find that the polarisation degree due to scattering in the (sub-)mm wavelength range is two to ten times lower in our simulations than what was found in observations associated with self-scattering.

    The optical depth and the anisotropy of the radiation field are the main parameters that control the polarisation degree due to self-scattering. The parameters of disk geometry and density distribution influence the polarisation degree only through one or both of these quantities. Therefore, the deduction of disk parameter values from only polarisation observations is highly degenerate. However, the transition from optically thick to optically thin can be traced by the radial distance of the maximum degree of polarisation if the dust albedo is high. Therefore, the radial density profile can be determined and compared to complementary, already existing estimates deduced from continuum intensity maps.
\begin{acknowledgement}
    This research was funded through the DFG grant WO 857/18-1.
\end{acknowledgement}




\bibliographystyle{aa}
\bibliography{lit}




\end{document}